\pdfoutput=1
\documentclass[12pt,a4paper]{article}
\usepackage{ifthen} % for conditional statements
\newboolean{pdflatex}
\setboolean{pdflatex}{true} % False for eps figures 

\newboolean{articletitles}
\setboolean{articletitles}{true} % False removes titles in references

\newboolean{uprightparticles}
\setboolean{uprightparticles}{true} %True for upright particle symbols

\def\paperauthors{LHCb collaboration} % Leave as is for PAPER and CONF
\def\paperasciititle{
Near-threshold DDbar spectroscopy and observation of a new charmonium state} % Set ASCII title here
\def\papertitle{Near\nobreakdash-threshold $\D\Dbar$~spectroscopy
and observation of a~new charmonium state} % Latex formatted title
\def\paperkeywords{{High Energy Physics}, {LHCb}} % Comma separated list
\def\papercopyright{\the\year\ CERN for the benefit of the LHCb collaboration} % new since 9/Apr/2018
\def\paperlicence{CC-BY-4.0 licence}
\def\paperlicenceurl{https://creativecommons.org/licenses/by/4.0/}

\usepackage{lscape}
\usepackage{graphpap} 
\usepackage{multirow} 
\usepackage{rotating} 
\usepackage{mathrsfs}
\usepackage{mathtools} %% dcases 
\usepackage{dashrule}
\usepackage{tikz}
\usetikzlibrary{patterns}
\usepackage{nicefrac} %% 1/2, ...
\usepackage{transparent}
\usepackage{afterpage} 
\usepackage[top=1in, bottom=1.25in, left=1in, right=1in]{geometry}

\usepackage{microtype}
\usepackage{lineno}  % for line numbering during review
\usepackage{xspace} % To avoid problems with missing or double spaces after
\usepackage{caption} %these three command get the figure and table captions automatically small

\usepackage{graphicx}  % to include figures (can also use other packages)
\usepackage{color}
\usepackage{colortbl}
\graphicspath{{./figs/}} % Make Latex search fig subdir for figures
\DeclareGraphicsExtensions{.pdf,.PDF,png,.PNG}

\usepackage{amsmath} % Adds a large collection of math symbols
\usepackage{amssymb}
\usepackage{amsfonts}
\usepackage{upgreek} % Adds in support for greek letters in roman typeset

\newcommand*\patchAmsMathEnvironmentForLineno[1]{%
\expandafter\let\csname old#1\expandafter\endcsname\csname #1\endcsname
\expandafter\let\csname oldend#1\expandafter\endcsname\csname
end#1\endcsname
 \renewenvironment{#1}%
   {\linenomath\csname old#1\endcsname}%
   {\csname oldend#1\endcsname\endlinenomath}%
}
\newcommand*\patchBothAmsMathEnvironmentsForLineno[1]{%
  \patchAmsMathEnvironmentForLineno{#1}%
  \patchAmsMathEnvironmentForLineno{#1*}%
}
\AtBeginDocument{%
\patchBothAmsMathEnvironmentsForLineno{equation}%
\patchBothAmsMathEnvironmentsForLineno{align}%
\patchBothAmsMathEnvironmentsForLineno{flalign}%
\patchBothAmsMathEnvironmentsForLineno{alignat}%
\patchBothAmsMathEnvironmentsForLineno{gather}%
\patchBothAmsMathEnvironmentsForLineno{multline}%
\patchBothAmsMathEnvironmentsForLineno{eqnarray}%
}

\usepackage{hyperxmp}

\usepackage[pdftex,
            pdfauthor={\paperauthors},
            pdftitle={\paperasciititle},
            pdfkeywords={\paperkeywords},
            pdfcopyright={Copyright (C) \papercopyright},
            pdflicenseurl={\paperlicenceurl}]{hyperref}

\usepackage[colorinlistoftodos,textsize=scriptsize]{todonotes}

\usepackage[all]{hypcap} % Internal hyperlinks to floats.

\usepackage{xspace} 
\usepackage{upgreek}

\def\lhcb   {\mbox{LHCb}\xspace}

\def\MagUp {\mbox{\em Mag\kern -0.05em Up}\xspace}

\ifthenelse{\boolean{uprightparticles}}%
{
 
 \def\Pgamma      {\ensuremath{\upgamma}\xspace}

 \def\Peta        {\ensuremath{\upeta}\xspace}

 \def\Pmu         {\ensuremath{\upmu}\xspace}

 \def\Ppi         {\ensuremath{\uppi}\xspace}

 \def\Pchi        {\ensuremath{\upchi}\xspace}                 
 \def\Ppsi        {\ensuremath{\uppsi}\xspace}                 
 \def\Pomega      {\ensuremath{\upomega}\xspace}                 

 \def\PDelta      {\ensuremath{\Delta}\xspace}                 
 \def\PXi         {\ensuremath{\Xi}\xspace}                 
 \def\PLambda     {\ensuremath{\Lambda}\xspace}                 
 \def\PSigma      {\ensuremath{\Sigma}\xspace}                 
 \def\POmega      {\ensuremath{\Omega}\xspace}                 
 \def\PUpsilon    {\ensuremath{\Upsilon}\xspace}

 \def\PB      {\ensuremath{\mathrm{B}}\xspace}                 
                  
 \def\PD      {\ensuremath{\mathrm{D}}\xspace}

 \def\PJ      {\ensuremath{\mathrm{J}}\xspace}                 
 \def\PK      {\ensuremath{\mathrm{K}}\xspace}

 \def\Pb      {\ensuremath{\mathrm{b}}\xspace}                 
 \def\Pc      {\ensuremath{\mathrm{c}}\xspace}

 \def\Pi      {\ensuremath{\mathrm{i}}\xspace}

 \def\Pp      {\ensuremath{\mathrm{p}}\xspace}

 \def\Ps      {\ensuremath{\mathrm{s}}\xspace}

 \def\thebaroffset{0.0em}
}
{
 
 \def\Pgamma      {\ensuremath{\gamma}\xspace}

 \def\Peta        {\ensuremath{\eta}\xspace}

 \def\Pmu         {\ensuremath{\mu}\xspace}

 \def\Ppi         {\ensuremath{\pi}\xspace}

 \def\Pchi        {\ensuremath{\chi}\xspace}                 
 \def\Ppsi        {\ensuremath{\psi}\xspace}                 
 \def\Pomega      {\ensuremath{\omega}\xspace}                 
 \mathchardef\PDelta="7101
 \mathchardef\PXi="7104
 \mathchardef\PLambda="7103
 \mathchardef\PSigma="7106
 \mathchardef\POmega="710A
 \mathchardef\PUpsilon="7107
                  
 \def\PB      {\ensuremath{B}\xspace}                 
                  
 \def\PD      {\ensuremath{D}\xspace}

 \def\PJ      {\ensuremath{J}\xspace}                 
 \def\PK      {\ensuremath{K}\xspace}

 \def\Pb      {\ensuremath{b}\xspace}                 
 \def\Pc      {\ensuremath{c}\xspace}

 \def\Pi      {\ensuremath{i}\xspace}

 \def\Pp      {\ensuremath{p}\xspace}

 \def\Ps      {\ensuremath{s}\xspace}

 \def\thebaroffset{0.18em}
}
\def\shortenleft{0.06em}
\def\shortenright{0.15em}
\newcommand{\offsetoverline}[2][\thebaroffset]{\kern #1\kern \shortenleft\overline{\kern -#1 \kern -\shortenleft #2\kern -\shortenright}\kern \shortenright}%

\makeatletter
\ifcase \@ptsize \relax% 10pt
  \newcommand{\miniscule}{\@setfontsize\miniscule{4}{5}}% \tiny: 5/6
\or% 11pt
  \newcommand{\miniscule}{\@setfontsize\miniscule{5}{6}}% \tiny: 6/7
\or% 12pt
  \newcommand{\miniscule}{\@setfontsize\miniscule{5}{6}}% \tiny: 6/7
\fi
\makeatother

\DeclareRobustCommand{\optbar}[1]{\shortstack{{\miniscule (\rule[.5ex]{1.25em}{.18mm})}
  \\ [-.7ex] $#1$}}

\def\mumu       {{\ensuremath{\Pmu^+\Pmu^-}}\xspace}

\def\g      {{\ensuremath{\Pgamma}}\xspace}

\def\squark    {{\ensuremath{\Ps}}\xspace}

\def\cquark    {{\ensuremath{\Pc}}\xspace}
\def\cquarkbar {{\ensuremath{\overline \cquark}}\xspace}

\def\bquark    {{\ensuremath{\Pb}}\xspace}

\def\pion   {{\ensuremath{\Ppi}}\xspace}
\def\piz    {{\ensuremath{\pion^0}}\xspace}
\def\pip    {{\ensuremath{\pion^+}}\xspace}
\def\pim    {{\ensuremath{\pion^-}}\xspace}

\def\kaon    {{\ensuremath{\PK}}\xspace}
\def\KorKbar {\kern \thebaroffset\optbar{\kern -\thebaroffset \PK}{}\xspace}

\def\Kp      {{\ensuremath{\kaon^+}}\xspace}
\def\Km      {{\ensuremath{\kaon^-}}\xspace}

\def\KS      {{\ensuremath{\kaon^0_{\mathrm{S}}}}\xspace}

\def\Dbar    {{\ensuremath{\offsetoverline{\PD}}}\xspace}
\def\D       {{\ensuremath{\PD}}\xspace}

\def\DorDbar {\kern \thebaroffset\optbar{\kern -\thebaroffset \PD}\xspace}
\def\Dz      {{\ensuremath{\D^0}}\xspace}
\def\Dzb     {{\ensuremath{\Dbar{}^0}}\xspace}
\def\Dp      {{\ensuremath{\D^+}}\xspace}
\def\Dm      {{\ensuremath{\D^-}}\xspace}

\def\Ds      {{\ensuremath{\D^+_\squark}}\xspace}

\def\Dsm     {{\ensuremath{\D^-_\squark}}\xspace}

\def\B       {{\ensuremath{\PB}}\xspace}

\def\BorBbar {\kern \thebaroffset\optbar{\kern -\thebaroffset \PB}\xspace}
\def\Bz      {{\ensuremath{\B^0}}\xspace}

\def\Bu      {{\ensuremath{\B^+}}\xspace}

\def\jpsi     {{\ensuremath{{\PJ\mskip -3mu/\mskip -2mu\Ppsi\mskip 2mu}}}\xspace}

\def\Y#1S{\ensuremath{\PUpsilon{(#1S)}}\xspace}

\def\proton      {{\ensuremath{\Pp}}\xspace}

\def\Lz          {{\ensuremath{\PLambda}}\xspace}

\def\LorLbar     {\kern \thebaroffset\optbar{\kern -\thebaroffset \PLambda}\xspace}

\def\Lc          {{\ensuremath{\Lz^+_\cquark}}\xspace}

\def\Lb           {{\ensuremath{\Lz^0_\bquark}}\xspace}

\newcommand{\decay}[2]{\mbox{\ensuremath{#1\!\to #2}}\xspace}         % {\Pa}{\Pb \Pc}

\def\to                 {\ensuremath{\rightarrow}\xspace}

\def\AT#1     {\ensuremath{A_{\mathrm{T}}^{#1}}\xspace}           % 2

\def\C#1      {\ensuremath{\mathcal{C}_{#1}}\xspace}                       % 9
\def\Cp#1     {\ensuremath{\mathcal{C}_{#1}^{'}}\xspace}                    % 7
\def\Ceff#1   {\ensuremath{\mathcal{C}_{#1}^{\mathrm{(eff)}}}\xspace}        % 9  
\def\Cpeff#1  {\ensuremath{\mathcal{C}_{#1}^{'\mathrm{(eff)}}}\xspace}       % 7
\def\Ope#1    {\ensuremath{\mathcal{O}_{#1}}\xspace}                       % 2
\def\Opep#1   {\ensuremath{\mathcal{O}_{#1}^{'}}\xspace}                    % 7

\newcommand{\nospaceunit}[1]{\ensuremath{\text{#1}}}       
\newcommand{\aunit}[1]{\ensuremath{\text{\,#1}}}       
\newcommand{\tev}{\aunit{Te\kern -0.1em V}\xspace}
\newcommand{\gev}{\aunit{Ge\kern -0.1em V}\xspace}
\newcommand{\mev}{\aunit{Me\kern -0.1em V}\xspace}
\newcommand{\kev}{\aunit{ke\kern -0.1em V}\xspace}
\newcommand{\ev}{\aunit{e\kern -0.1em V}\xspace}
\newcommand{\mevc}{\ensuremath{\aunit{Me\kern -0.1em V\!/}c}\xspace}
\newcommand{\gevc}{\ensuremath{\aunit{Ge\kern -0.1em V\!/}c}\xspace}
\newcommand{\mevcc}{\ensuremath{\aunit{Me\kern -0.1em V\!/}c^2}\xspace}
\newcommand{\gevcc}{\ensuremath{\aunit{Ge\kern -0.1em V\!/}c^2}\xspace}
\newcommand{\kevcc}{\ensuremath{\aunit{ke\kern -0.1em V\!/}c^2}\xspace}
\def\mum  {\ensuremath{\,\upmu\nospaceunit{m}}\xspace}

\def\fb   {\ensuremath{\aunit{fb}}\xspace}
\def\invfb   {\ensuremath{\fb^{-1}}\xspace}

\newcommand{\chisq}{\ensuremath{\chi^2}\xspace}

\def\gsim{{~\raise.15em\hbox{$>$}\kern-.85em
          \lower.35em\hbox{$\sim$}~}\xspace}
\def\lsim{{~\raise.15em\hbox{$<$}\kern-.85em
          \lower.35em\hbox{$\sim$}~}\xspace}

\def\pt         {\ensuremath{p_{\mathrm{T}}}\xspace}

\def\ptot       {\ensuremath{p}\xspace}

\def\evtgen     {\mbox{\textsc{EvtGen}}\xspace}

\def\geant      {\mbox{\textsc{Geant4}}\xspace}

\def\photos     {\mbox{\textsc{Photos}}\xspace}

\def\pythia     {\mbox{\textsc{Pythia}}\xspace}

\xspace

\def\tell1  {TELL1\xspace}
\def\ukl1   {UKL1\xspace}

\usepackage{cite} % Allows for ranges in citations
\usepackage{mciteplus}

\usepackage{longtable} % only for template; not usually to be used in PAPERs

\begin{document}

%%%%%%%%%%%%%%%%%%%%%%%%%
%%%%% Title     %%%%%%%%%
%%%%%%%%%%%%%%%%%%%%%%%%%
\renewcommand{\thefootnote}{\fnsymbol{footnote}}
\setcounter{footnote}{1}

% %%%%%%% CHOOSE TITLE PAGE--------
%\onecolumn
%\input{title-LHCb-INT}
%\input{title-LHCb-ANA}
%\input{title-LHCb-CONF}
% $Id: title-LHCb-PAPER.tex 122889 2018-08-17 17:59:55Z pkoppenb $
% ===============================================================================
% Purpose: LHCb-PAPER journal paper title page template
% Author: 
% Created on: 2010-09-25
% ===============================================================================

%%%%%%%%%%%%%%%%%%%%%%%%%
%%%%%  TITLE PAGE  %%%%%%
%%%%%%%%%%%%%%%%%%%%%%%%%
\begin{titlepage}
\pagenumbering{roman}

% Header ---------------------------------------------------
\vspace*{-1.5cm}
\centerline{\large EUROPEAN ORGANIZATION FOR NUCLEAR RESEARCH (CERN)}
\vspace*{1.5cm}
\noindent
\begin{tabular*}{\linewidth}{lc@{\extracolsep{\fill}}r@{\extracolsep{0pt}}}
\ifthenelse{\boolean{pdflatex}}% Logo format choice
{\vspace*{-1.5cm}\mbox{\!\!\!\includegraphics[width=.14\textwidth]{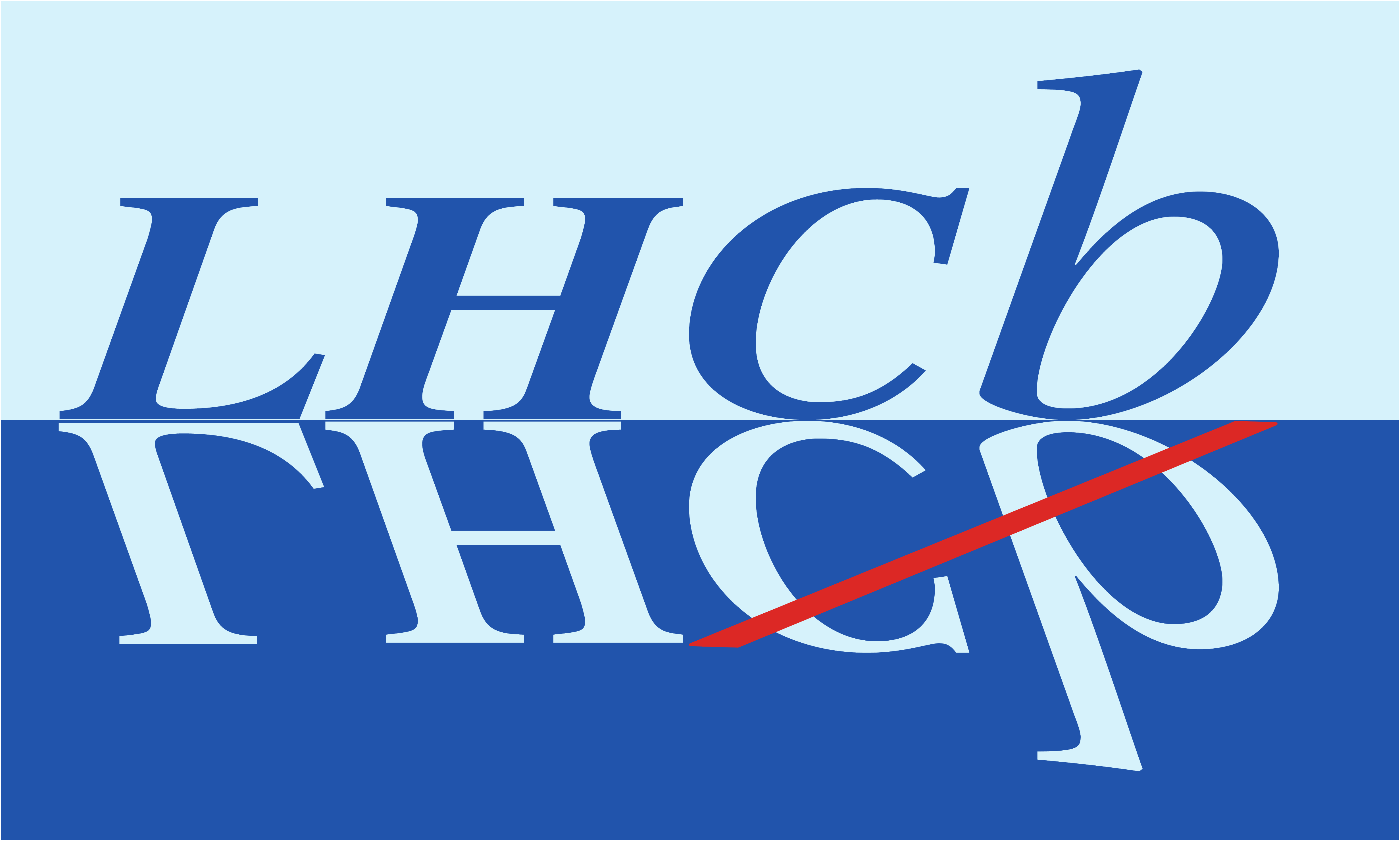}} & &}%
{\vspace*{-1.2cm}\mbox{\!\!\!\includegraphics[width=.12\textwidth]{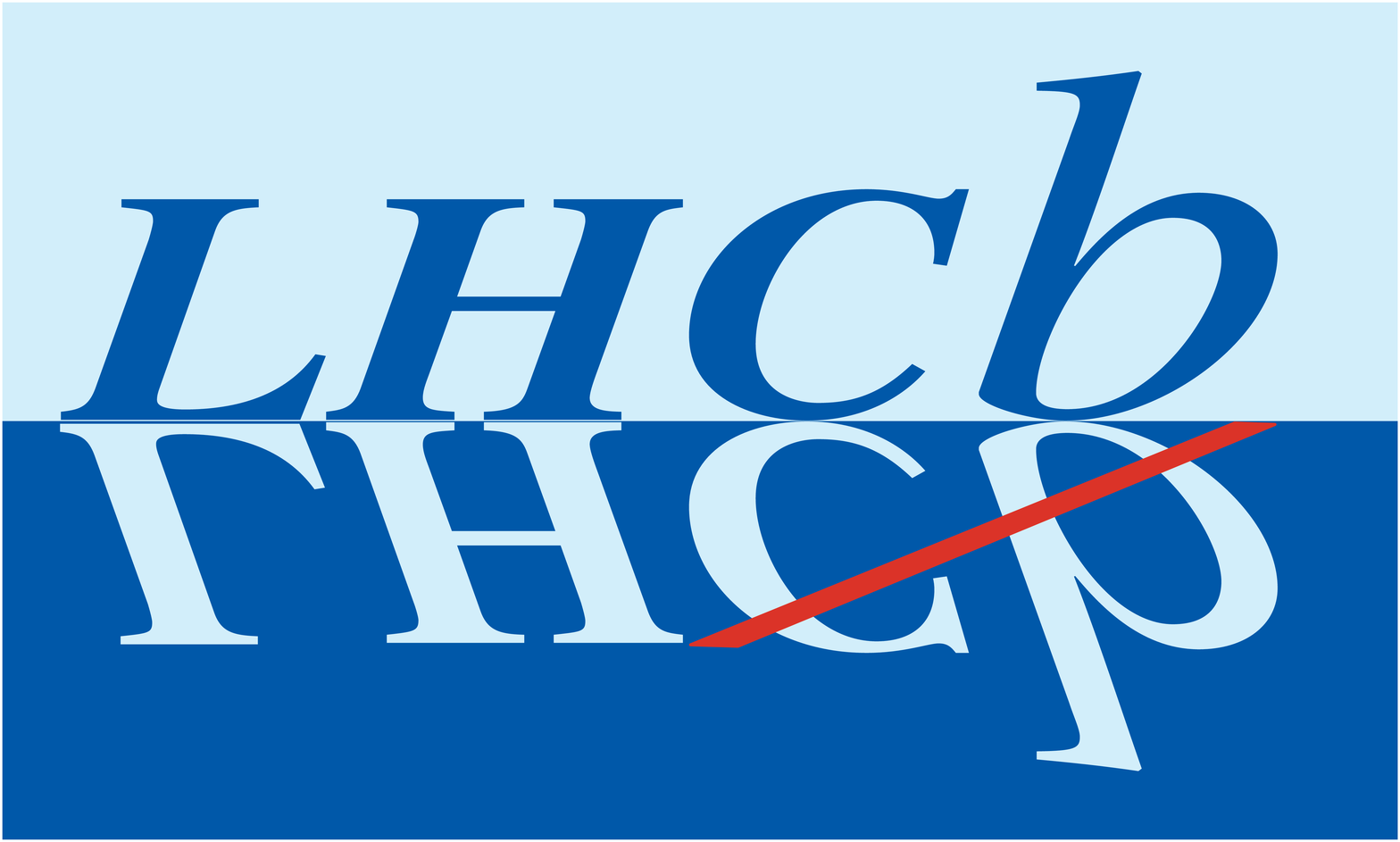}} & &}%
\\
 & & CERN-EP-2019-047 \\  % ID 
 & & LHCb-PAPER-2019-005 \\  % ID 
 %% & & \today \\ % Date - Can also hardwire e.g.: 23 March 2010
 & & March 28, 2019 \\ % Date - Can also hardwire e.g.: 23 March 2010
 %% & & v1.5.0\\ 
 & & \\
% not in paper \hline
\end{tabular*}

\vspace*{1.5cm}

% Title --------------------------------------------------
{\normalfont\bfseries\boldmath\huge
\begin{center}
% DO NOT EDIT HERE. Instead edit macro in main.tex to keep metadata correct
  \papertitle 
\end{center}
}

\vspace*{1.0cm}

% Authors -------------------------------------------------
\begin{center}
%In the footnote, replace 'paper' by 'Letter' in case of submission to PRL or PLB 
% Edit macro in main.tex to keep metadata correct
\paperauthors\footnote{Authors are listed at the end of this paper.}
\end{center}

\vspace{\fill}

% Abstract -----------------------------------------------
\begin{abstract}
  \noindent
Using proton\nobreakdash-proton collision
data, corresponding to an~integrated luminosity of 9\invfb, 
collected with the~LHCb detector between 2011 and 2018,
a~new narrow charmonium state, the~$\mathrm{X(3842)}$~resonance, 
is observed in the~decay modes $\decay{\mathrm{X(3842)}}{\Dz\Dzb}$ and 
$\decay{\mathrm{X(3842)}}{\Dp\Dm}$. %% with overwhelming significance. 
The~mass and the~natural width 
of this state are measured to be 
\begin{eqnarray*}
  m_{\mathrm{X(3842)}}  & = & 3842.71           \pm 0.16 \pm 0.12 \mevcc\,, \\
  \Gamma_{\mathrm{X(3842)}} & = & \phantom{000}2.79 \pm 0.51 \pm 0.35  \mev\,,
\end{eqnarray*}
where the~first uncertainty is statistical and the~second is systematic.
The~observed mass and narrow natural width suggest the~interpretation of the~new state 
as the~unobserved spin\nobreakdash-3 $\Ppsi_3\mathrm{\left(1^3D_3\right)}$~charmonium state.

In addition, prompt hadroproduction of the
$\Ppsi\mathrm{(3770)}$~and
$\Pchi_{\cquark2}\mathrm{(3930)}$~states is observed for the~first time, and the~parameters of these~states are measured to be 
\begin{eqnarray*}
  %%m_{\Ppsi\mathrm{(3770)}}  & = & 3778.13 \pm 0.70 \pm 0.63 \mevcc\,, \\
  %% m_{\Pchi_{\cquark2}\mathrm{(3930)}}  & = & 3921.90 \pm 0.55 \pm 0.19 \mevcc\,, \\ 
  %% \Gamma_{\Pchi_{\cquark2}\mathrm{(3930)}} & = &\phantom{00}36.64 \pm 1.88 \pm 0.85 \mev\,, 
  m_{\Ppsi\mathrm{(3770)}}  & = & 3778.1 \pm 0.7 \pm 0.6 \mevcc\,, \\
  m_{\Pchi_{\cquark2}\mathrm{(3930)}}  & = & 3921.9 \pm 0.6 \pm 0.2 \mevcc\,, \\ 
  \Gamma_{\Pchi_{\cquark2}\mathrm{(3930)}} & = &\phantom{00}36.6 \pm 1.9 \pm 0.9 \mev\,, 
\end{eqnarray*}
where the~first uncertainty is statistical and the~second is systematic.

\end{abstract}

\vspace*{1.0cm}

\begin{center}
  Published in \href{https://doi.org/10.1007/JHEP07(2019)035}{JHEP 1907 (2019) 035}
\end{center}

\vspace{\fill}

{\footnotesize 
% Edit macro in main.tex to keep metadata correct
\centerline{\copyright~\papercopyright. \href{\paperlicenceurl}{\paperlicence}.}}
\vspace*{2mm}

\end{titlepage}

%%%%%%%%%%%%%%%%%%%%%%%%%%%%%%%%
%%%%%  EOD OF TITLE PAGE  %%%%%%
%%%%%%%%%%%%%%%%%%%%%%%%%%%%%%%%

%  empty page follows the title page ----
\newpage
\setcounter{page}{2}
\mbox{~}
%\newpage
%
%% Author List ----------------------------
%%  You need to get a new author list!
%\input{LHCb_authorlist.tex}
%
%The author list for journal publications is provided by the Membership Committee shortly after 'approval to go to paper' has been given.
%%It will be made available on the page
%%\verb!http://www.physik.uzh.ch/~strauman/forMemCo/LHCb-PAPER-XXXX-XXX/! .
%It will be sent to you by email shortly after a paper number has beens assigned.
%The author list should be included already at first circulation, 
%to allow new members of the collaboration to verify whether they have been included correctly.
%Occasionally a misspelled name is corrected or associated institutions become full members.
%In that case, a new author list will be sent to you.
%In case line numbering doesn't work well after including the authorlist, try moving the \verb!\bigskip! after the last author to a separate line.
%
%
%The authorship for Conference Reports should be ``The LHCb
%  collaboration'', with a footnote giving the name(s) of the contact
%  author(s), but without the full list of collaboration names.

\cleardoublepage

%\twocolumn
% %%%%%%%%%%%%% ---------

\renewcommand{\thefootnote}{\arabic{footnote}}
\setcounter{footnote}{0}

%%%%%%%%%%%%%%%%%%%%%%%%%%%%%%%%
%%%%%  Table of Content   %%%%%%
%%%%%%%%%%%%%%%%%%%%%%%%%%%%%%%%
%%%% Uncomment next 2 lines if desired
%\tableofcontents
%\cleardoublepage

%%%%%%%%%%%%%%%%%%%%%%%%%
%%%%% Main text %%%%%%%%%
%%%%%%%%%%%%%%%%%%%%%%%%%

\pagestyle{plain} % restore page numbers for the main text
\setcounter{page}{1}
\pagenumbering{arabic}

%% Uncomment during review phase. 
%% Comment before a final submission.
%% \linenumbers

% You can include short sections directly in the main tex file.
% However, for larger papers it is desirable to split the text into
% several semiautonomous files, which can be revised independently.
% This is especially useful when developing a document in
% collaboration with several people, since then different parts can be
% edited independently.  This type of file organization is shown here.
% 

\section{Introduction}\label{sec:intro} 
Since the~discovery of the~$\jpsi$~resonance in 1974~\cite{PhysRevLett.33.1404,PhysRevLett.33.1406},
the~spectrum of hidden charm mesons has been mapped out experimentally
with high precision. %\cite{Eichten:2007qx}.
Theoretically, the~spectra and properties of
these states are well described by potential models~\cite{Eichten:1978tg}.
In~recent years, 
there has been a~revival of interest in charmonium spectroscopy initially triggered by
the~discovery of the~$\Pchi_{\cquark1}\mathrm{(3872)}$ meson\footnote{Also known as the~$\mathrm{X(3872)}$~state.} by 
the~Belle experiment~\cite{Choi:2003ue} and the~subsequent observation of other states that do not fit into the~conventional
hidden\nobreakdash-charm spectrum. 
To~be confident that the~new states are exotic in nature, all predicted 
$\cquark\cquarkbar$~states need to be accounted for.

Amongst the~expected charmonia close to $\D\Dbar$~threshold,
%% two states with quantum numbers~
the~states $\Peta_{\cquark2}\mathrm{(1^1D_2)}$ and 
$\Ppsi_3\mathrm{(1^3D_3)}$ remain undiscovered~\cite{Eichten:2002qv,Eichten:2005ga}. 
Though the~latter state lies above the open charm threshold, 
the %% ~Zweig\nobreakdash-allowed 
decay to the~$\D\Dbar$ final 
state is suppressed due to the~F\nobreakdash-wave centrifugal
barrier factor. Consequently, 
the~$\Ppsi_3\mathrm{(1^3D_3)}$ state is expected to be narrow with a~natural width of $1$--$2\mev$~\cite{Barnes:2003vb, Barnes:2005pb}. 
Predictions for the~mass of this state lie in the~range 
\mbox{$3815$--$3863\mevcc$}~\mbox{\cite{Radford:2007vd,Eichten:2005ga,Godfrey:1985xj,Eichten:1980mw, Fulcher:1991dm,Gupta:1986xt,Ebert:2002pp,Zeng:1994vj}}. 
Since it has negative 
C~parity,
it cannot be produced in either $\g\g$~annihilation or $\mathrm{gg}$~fusion.
In~Ref.~\cite{Barnes:2005pb} it is suggested that a~possible production mechanism for this state is via electric\nobreakdash-dipole radiative transitions from the~$\Pchi_{\cquark2}\mathrm{(2^3P_2)}$~tensor state.%~\cite{Barnes:2005pb}. 

In this paper,
the~observation of a new $\cquark\bar{\cquark}$~meson decaying to both
the $\Dp\Dm$~and $\Dz\Dzb$~final states is reported. 
The~data sample used for this analysis corresponds to an~integrated
luminosity of 9\invfb recorded with the~LHCb detector in $\proton\proton$~collisions
at centre\nobreakdash-of\nobreakdash-mass energies of 7, 8 and 13\tev,
during the~years 2011--2018.
The~mass and width of the~new state are quite similar 
to those expected for
the~missing $\Ppsi_3\mathrm{(1^3D_3)}$~state with $\mathrm{J}^{\mathrm{PC}}=3^{--}$.
In~addition, the~production of both $\Ppsi\mathrm{(3770)}$~and
$\Pchi_{\cquark2}\mathrm{(3930)}$ mesons~is observed.
The~first state is well known through measurements at $\mathrm{e}^+\mathrm{e}^-$~colliders,
%%
%% but so far has not been seen in a~hadronic environment. 
but so far it has only been observed in a~hadronic environment 
%% in the~$\decay{\Ppsi\mathrm{(3770)}}{\mumu}$~final state in %% the~$\decay{\Bu}{\Kp\mumu}$~decays~\cite{LHCb-PAPER-2013-039}.
in the~$\mumu$~mass spectrum of~$\decay{\Bu}{\Kp\mumu}$~decays\footnote{The~inclusion of charge\nobreakdash-conjugate processes is implied throughout the~paper.}~\cite{LHCb-PAPER-2013-039}.
The~latter state has only been previously observed 
in the~$\decay{\g\g}{\D\Dbar}$~process by the~Belle and BaBar experiments~\cite{Uehara:2005qd,Aubert:2010ab}. 
Both analyses prefer 
a~spin assignment of 2 for this state based upon one-dimensional angular distributions.

\section{The LHCb detector and simulation}
\label{sec:Detector}
The \lhcb detector~\cite{Alves:2008zz,LHCb-DP-2014-002} is a single-arm forward
spectrometer covering the~\mbox{pseudorapidity} range \mbox{$2<\eta <5$},
designed for the~study of particles containing \bquark or \cquark~quarks.
The detector includes a~high\nobreakdash-precision tracking system
consisting of a~silicon\nobreakdash-strip vertex detector surrounding
the~$\proton\proton$~interaction region~\cite{LHCb-DP-2014-001},
a~large\nobreakdash-area silicon\nobreakdash-strip detector located
upstream of a~dipole magnet with a~bending power of about~$4{\mathrm{\,Tm}}$,
and three stations of silicon\nobreakdash-strip detectors and straw
drift tubes~\cite{LHCb-DP-2013-003,LHCb-DP-2017-001} placed downstream of the magnet.
The~tracking system provides a~measurement of the~momentum, \ptot, of charged particles with
a~relative uncertainty that varies from~0.5\% at low momentum to~1.0\% at~200\gevc. The
momentum scale is calibrated using samples of $\decay{\jpsi}{\mumu}$ 
and $\decay{\Bu}{\jpsi\Kp}$~decays collected concurrently
with the~data sample used for this analysis~\cite{LHCb-PAPER-2012-048,LHCb-PAPER-2013-011}. 
The~relative accuracy of this
procedure is estimated to be $3 \times 10^{-4}$ using samples of other
fully reconstructed $\bquark$~hadrons, $\PUpsilon$~and
$\KS$~mesons. The~minimum distance of a~track to
a~primary vertex\,(PV), the~impact parameter (IP), 
is measured with a~resolution of~$(15+29/\pt)\mum$,
where \pt is the~component of the~momentum transverse to the~beam, in\,\gevc.
Different types of charged hadrons are distinguished using information
from two ring\nobreakdash-imaging Cherenkov detectors\,(RICH)~\cite{LHCb-DP-2012-003}. 
Photons, electrons and hadrons are identified by a~calorimeter system consisting of
scintillating\nobreakdash-pad and preshower detectors,
an~electromagnetic
and a~hadronic calorimeter.
Muons~are identified by a~system composed of alternating layers of iron and multiwire
proportional chambers~\cite{LHCb-DP-2012-002}.

The~online event selection is performed by a~trigger~\cite{LHCb-DP-2012-004}, 
which consists of a~hardware stage, based on information from the~calorimeter and muon
systems, followed by a~software stage, which applies a~full event
reconstruction.
At the~hardware trigger stage, events are required to have a~muon with high \pt or
a~hadron, photon or electron with high transverse energy in the~calorimeters.
%% For~hadrons,
%% the~transverse energy threshold is~3.5\gev.
The~software trigger requires a~two-, three- or four\nobreakdash-track
secondary vertex with a~significant displacement from any primary
$\proton\proton$~interaction vertex.
At~least one charged particle
must have transverse momentum~\mbox{$\pt > 1.6\gevc$} and be
inconsistent with originating from a~PV.

%%% 
The~analysis procedure is validated using 
a~simulation in which $\proton\proton$~collisions 
are generated using
\pythia~\cite{Sjostrand:2007gs,*Sjostrand:2006za}
with a~specific \lhcb
configuration~\cite{LHCb-PROC-2010-056}.  Decays of unstable particles
are described by \evtgen~\cite{Lange:2001uf}, in which final-state
radiation is generated using \photos~\cite{Golonka:2005pn}. 
The~interaction of the~generated particles with the~detector, 
and its response, are implemented using the~\geant
toolkit~\cite{Allison:2006ve, *Agostinelli:2002hh} as described in
Ref.~\cite{LHCb-PROC-2011-006}.
%% \clearpage

\section{Selection}
\label{sec:selection}

The criteria used to select \Dz and \Dp~candidates are similar to those described in Refs.~\cite{LHCb-PAPER-2012-003,LHCb-PAPER-2013-062,LHCb-PAPER-2015-046}. 
The~selection starts from  good\nobreakdash-quality charged tracks with 
$\pt>250\mevc$ that are inconsistent with being 
produced in a~$\proton\proton$~interaction vertex.  
Selected tracks are required to be identified 
as either kaons or pions using information from the~RICH detectors, 
and are then used to build \Dz and \Dp~candidates reconstructed in  
the~$\decay{\Dz}{\Km\pip}$ and $\decay{\Dp}{\Km\pip\pip}$~decay modes. 
The~tracks forming \Dz and \Dp~candidates are required 
to originate from a~common vertex. 
To~reduce combinatorial background, 
the~decay time of \Dz~and~\Dp~candidates 
is required to exceed~$100\mum/c$ and the~momentum 
direction to be consistent with the~vector from 
the~primary to the~secondary vertex.
The~latter requirement also reduces the~contribution from charm hadrons produced in the~weak  decays of long\nobreakdash-lived beauty hadrons.
%%
%% Full~decay chain fits are applied for \Dz and \Dp candidates~\cite{Hulsbergen:2005pu}.
Selected \Dz and \Dp~candidates, generically referred to as  \D~candidates hereafter, with \mbox{$\pt>1\gevc$}
are combined to form $\Dz\Dzb$ and $\Dp\Dm$~candidates.
A~fit is performed for each $\D\Dbar$~candidate~\cite{Hulsbergen:2005pu}, such that  both~\D~mesons are required to originate from a~common vertex 
that is consistent with the~PV~location. 
A~requirement on the~fit \chisq 
reduces, to a~negligible level, the~background from 
$\D$ and $\Dbar$~candidates produced in
%the~pile\nobreakdash-up of 
two independent $\proton\proton$~interactions,
and further suppresses the~contribution from beauty hadrons.  

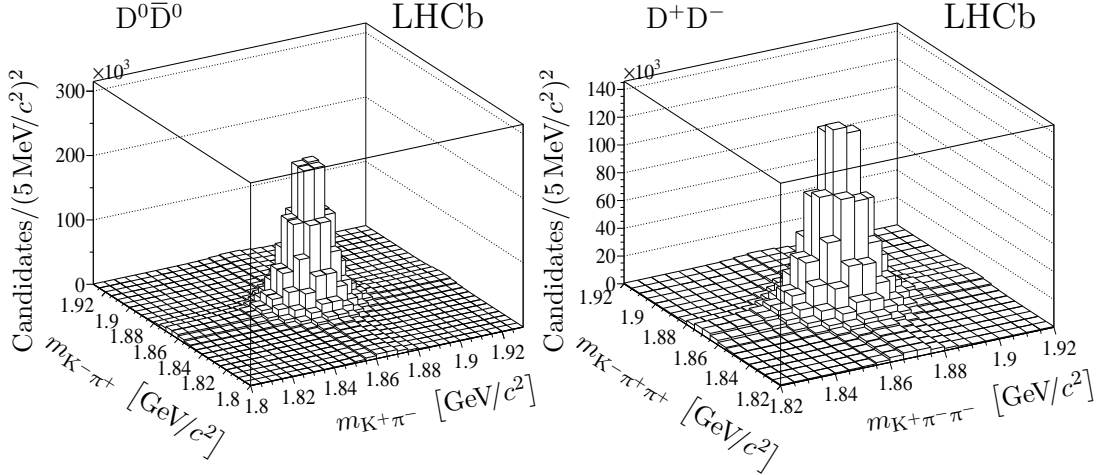
\begin{figure}[t]
  \setlength{\unitlength}{1mm}
  \centering
  \begin{picture}(150,65)
    %% 
    %%\graphpaper[5](-10,-10)(170,85)
    %% 
    % 
    \put(  0,  0){ 
      \includegraphics*[width=75mm,height=65mm,%
      %% ]{D0D0_lego.pdf}
      ]{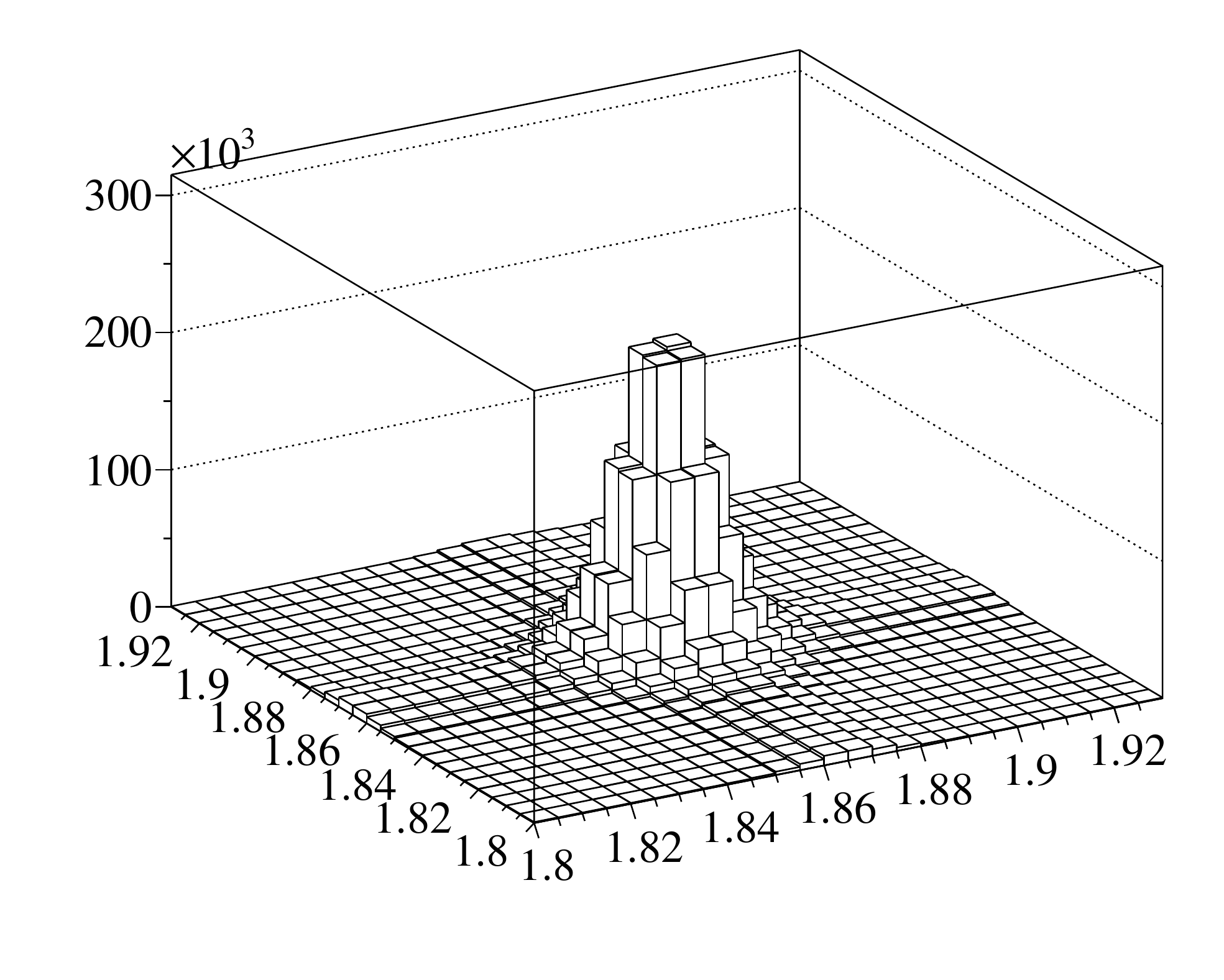}
    }
    \put( 75,  0){ 
      \includegraphics*[width=75mm,height=65mm,%
      %% ]{DpDm_lego.pdf}
      ]{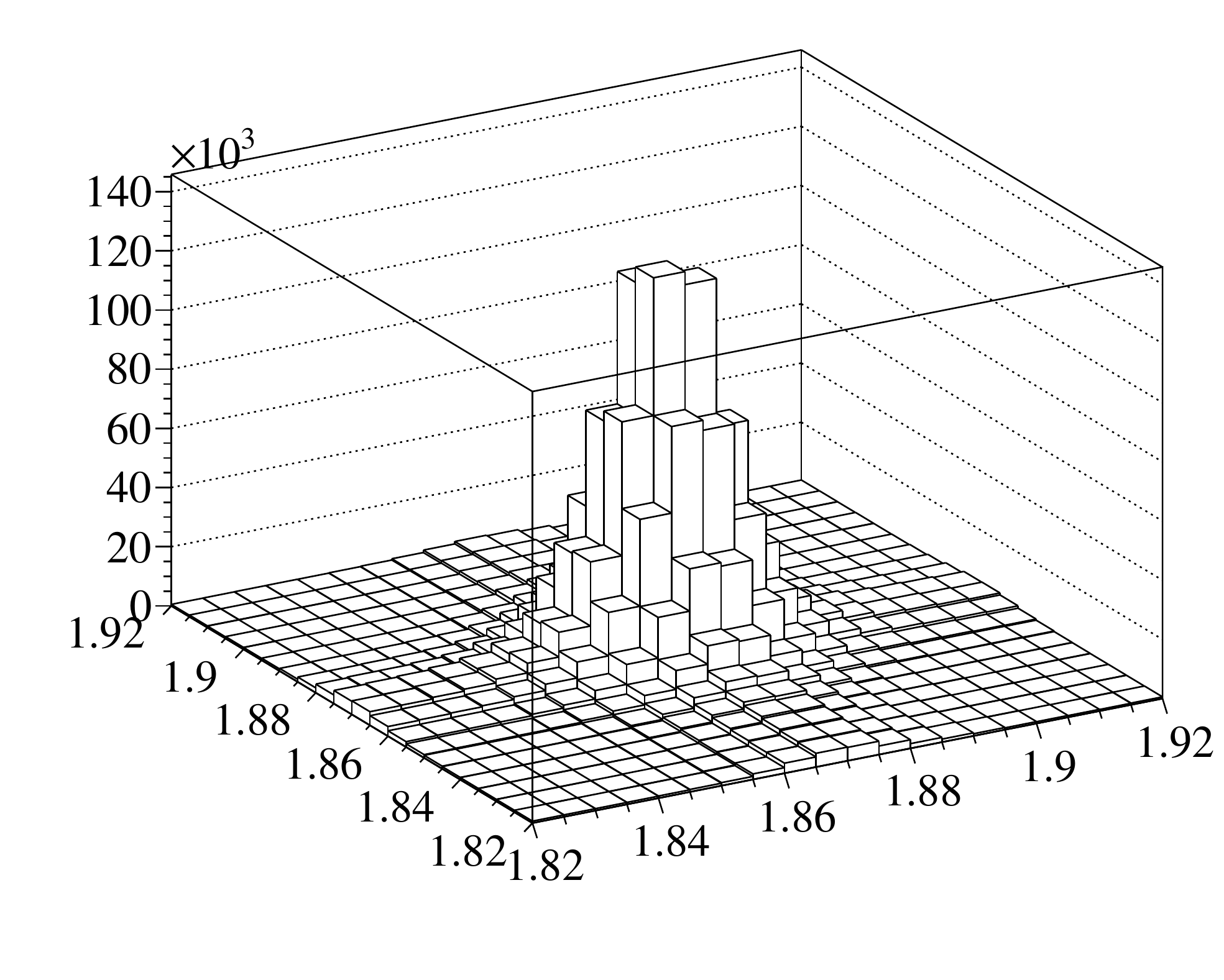}
    }
    \put( 3.5,16)  { 
      \begin{rotate}{-32} \small % \tiny
        $m_{\Km\pip}~\left[\!\gevcc\right]$
      \end{rotate}
    }
    \put(78.5,16)  { 
      \begin{rotate}{-32} \small % \tiny
        $m_{\Km\pip\pip}~\left[\!\gevcc\right]$
      \end{rotate}
    }
    \put( 45, 4.0)  { 
      \begin{rotate}{10}  \small % \tiny
        $m_{\Kp\pim}~\left[\!\gevcc\right]$
      \end{rotate}
    }
    \put(120, 4.0)  { 
      \begin{rotate}{10} \small % \tiny
        $m_{\Kp\pim\pim}~\left[\!\gevcc\right]$
      \end{rotate}
    }
    \put(  0, 15) {\small\begin{sideways}Candidates/$(5\mevcc)^2$\end{sideways}} 
    \put( 75, 15) {\small\begin{sideways}Candidates/$(5\mevcc)^2$\end{sideways}} 
    %% 
    %% \put( 60,55){\small{LHCb}}
    %% \put(135,55){\small{LHCb}}
    %% \put( 30,  0) { $m(\Km\pip\pip)~~~\left[\!\gevcc\right]$} 
    %% \put(105,  0) { $m(\Kp\pim\pim)~~~\left[\!\gevcc\right]$} 
    %
    %%\put( 5,62) {\small$\begin{array}{l}\mathrm{LHCb} \\ 
    %% \sqs=7,8\&13\tev \\ 9\invfb \end{array}$}
    \put(15,61){$\Dz\Dzb$}
    \put(90,61){$\Dp\Dm$}
    \put( 54,61){\large{LHCb}}
    \put(132,61){\large{LHCb}}
    %%\put( 55,61){LHCb}
    %%\put(130,61){LHCb}
    %%
  \end{picture}
  \caption { \small
    Distributions of
    (left)~$m_{\Km\pip}$ versus $m_{\Kp\pim}$ and  
    (right)~$m_{\Km\pip\pip}$ versus $m_{\Kp\pim\pim}$ 
    for selected $\D\Dbar$~candidates.
  }
  \label{fig:dd_lego}
\end{figure}

The~two\nobreakdash-dimensional distributions for 
the~\D~and $\Dbar$~masses are shown in~Fig.~\ref{fig:dd_lego}. 
Only \D~candidates with mass 
within $\pm20\mevcc$\,(approximately~$\pm3\sigma$) of the~known 
$\D$\nobreakdash-meson masses~\cite{PDG2018} are kept for subsequent analysis.
The~purity of the~selected samples is 
88\% and 83\% for the~$\Dz\Dzb$ and $\Dp\Dm$~modes, respectively.

%%\clearpage

\vspace{1cm}

\section{ {\boldmath{$\D\Dbar$}}~mass spectra}
\label{sec:spectra}
\begin{figure}[tb]
  \setlength{\unitlength}{1mm}
  \centering
  \begin{picture}(150,120)
    %% 
    %% \graphpaper[5](-10,-10)(170,140)
    %% 
    %
    \put(  0,  0){ 
      \includegraphics*[width=150mm,height=120mm,%
      %% ]{DD_mass_hatched.pdf}
      ]{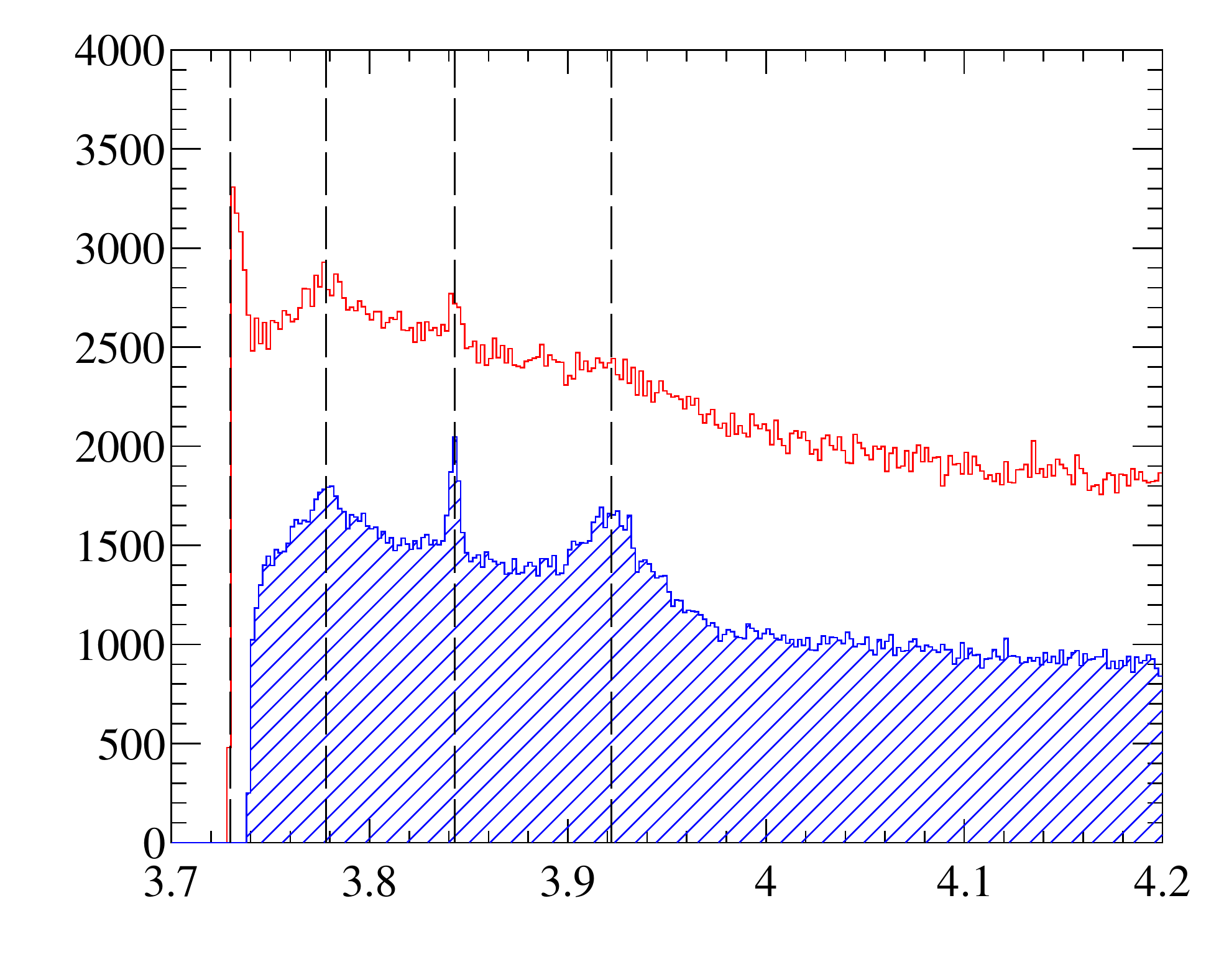}
    }
    \put(  0, 65) {\large\begin{sideways}Candidates/(2\mevcc)\end{sideways}} 
    \put( 70,  2) {\LARGE$m_{\D\Dbar}$} \put(119, 2) {\LARGE$\left[\!\gevcc\right]$} 
    \put(78,105) {\begin{tikzpicture}[x=1mm,y=1mm]\draw[thin,red]  (0,0) rectangle (10,2.0);\end{tikzpicture} }
    \put(78,100) {\begin{tikzpicture}[x=1mm,y=1mm]\draw[thin,blue,pattern=north east lines, pattern color=blue]  (0,0) rectangle (10,2.0);\end{tikzpicture} }
    \put(90,105) {\large$\Dz\Dzb$ } 
    \put(90,100) {\large$\Dp\Dm$  }
    \put(120,105) {\large{LHCb}}
    %%\put(107,105) {LHCb preliminary}
    %%\put(107,102) {\small$\begin{array}{r}\mathrm{LHCb} \\ \sqs=7,8\&13\tev \\ 9\invfb \end{array}$} 
    %%
    %% for Bolek
    %% \put(15,107){\small\fcolorbox{red}{yellow}{\color[rgb]{1,0,0}{\tiny{$\decay{\Pchi_{\cquark1}\mathrm{(3872)}}{\D{}^{\ast0}\Dzb}$}}}}
    %%
    %% \put(32,100){\small\fcolorbox{red}{yellow}{\color[rgb]{1,0,0}{\tiny{$\decay{\Ppsi\mathrm{(3770)}}{\D\Dbar}$}}}}
    %% 
    %% \put(49, 93){\small\fcolorbox{red}{yellow}{\color[rgb]{1,0,0}{\tiny{$\decay{\mathrm{X(3842)}}{\D\Dbar}$}}}}
    %% 
    %% \put(66, 86){\small\fcolorbox{red}{yellow}{\color[rgb]{1,0,0}{\tiny{$\decay{\Pchi_{\cquark2}\mathrm{(3930)}}{\D\Dbar}$}}}}
    %%
  \end{picture}
  \caption { \small
   The~mass spectra for selected $\D\Dbar$ combinations.
   The~open red histogram corresponds to $\Dz\Dzb$~pairs, 
   while the~hatched blue histogram corresponds to $\Dp\Dm$~pairs.
   Vertical black dashed lines help to identify the~peaks 
   from\,(left to right)
   $\decay{\Pchi_{\cquark1}\mathrm{(3872)}}{\D^{*0}\Dzb}$,
   $\decay{\Ppsi\mathrm{(3770)}}{\D\Dbar}$,
   $\decay{\mathrm{X(3842)}}{\D\Dbar}$ and  
   $\decay{\Pchi_{\cquark2}\mathrm{(3930)}}{\D\Dbar}$~decays.
  }
  \label{fig:dd_mass_wide}
\end{figure}

To~improve the~$\D\Dbar$~mass resolution,
a~new fit~\cite{Hulsbergen:2005pu} is performed
with the~masses of both~\D~candidates 
constrained to the~known values~\cite{PDG2018}.
After this fit, 
the~$\D\Dbar$~mass spectra for selected $\Dz\Dzb$ and $\Dp\Dm$~pairs 
close to the~$\D\Dbar$~threshold with \mbox{$m_{\D\Dbar}<4.2\gevcc$} 
are shown in Fig.~\ref{fig:dd_mass_wide}.
Four~peaking structures are seen:
\begin{itemize}
\item[-] 
A~narrow peak in the~$\Dz\Dzb$ spectrum just above the~threshold, interpreted as 
the~$\decay{\Pchi_{\cquark1}\mathrm{(3872)}}{\D^{*0}\Dzb}$~decay, 
followed  by 
$\decay{\D{}^{\ast0}}{\Dz\piz}$ or $\decay{\D{}^{\ast0}}{\Dz\g}$ ---
due to the~small energy release in this  decay,  
the~mass of the $\D\Dbar$~pair gives a~narrow peak in 
the~$\Dz\Dzb$~mass spectrum at the~$\Dz\Dzb$~threshold;
\item[-] A~broad peak close to $3780\mevcc$, visible both in $\Dz\Dzb$~and $\Dp\Dm$~mass spectra and associated with the~contribution from $\decay{\Ppsi\mathrm{(3770)}}{\D\Dbar}$~decays; 
\item[-] A~very narrow peak  at $m_{\D\Dbar}\approx3840\mevcc$, referred to hereafter as $\mathrm{X(3842)}$;
\item[-] A~wide structure in the~$\Dp\Dm$~mass spectrum at $m_{\Dp\Dm}\approx3920\mevcc$ also 
visible in the~$\Dz\Dzb$~mass spectrum and interpreted to be due to  $\decay{\Pchi_{\cquark2}\mathrm{(3930)}}{\D\Dbar}$~decays. 
\end{itemize}
To better parameterise the~background, 
fits to the~$\D\Dbar$~mass spectra are performed separately 
in three different overlapping mass regions:
a~narrow region \mbox{$3.80<m_{\D\Dbar}<3.88\gevcc$}
around the~$\mathrm{X(3842)}$~peak; 
the~high\nobreakdash-mass region \mbox{$3.8<m_{\D\Dbar}<4.2\gevcc$}
and  the~near\nobreakdash-threshold  region~\mbox{$m_{\D\Dbar}<3.88\gevcc$}.

%\vspace*{1cm}

\subsection{Mass region~{\boldmath{$3.80<m_{\D\Dbar}<3.88\gevcc$}}}\label{sec:narrow}

The~narrow natural width and the~mass of the $\mathrm{X(3842)}$~state suggest 
the~interpretation of the~$\mathrm{X(3842)}$~state as the~$\Ppsi_3\mathrm{\left(1^3D_3\right)}$~charmonium 
state with $\mathrm{J^{PC}=3^{--}}$~\cite{Barnes:2005pb}.
The~$\mathrm{X(3842)}$~signal is modelled by a relativistic Breit\nobreakdash--Wigner function with \mbox{Blatt}\nobreakdash--Weisskopf form factors~\cite{Blatt:1952ije}.
%%and a~meson radius parameter of~$3.5/\!\gev$. 
The~orbital angular momentum between the~$\D$ and $\Dbar$~mesons is assumed to be~$L=3$. 
Alternative hypotheses for the~spin assignment are discussed in Sect.~\ref{sec:systematics}.
The~relativistic Breit\nobreakdash--Wigner function is convolved with the~detector resolution, 
described by 
a~sum of two Gaussian functions with common mean
and parameters fixed from simulation. 
%% Three~resolution models are  found to describe simulated data well:
%% a~double\nobreakdash-Gaussian function, 
%% a~symmetric double\nobreakdash-sided Crystal Ball 
%% function~\cite{Skwarnicki:1986xj,LHCB-Paper-2011-013} 
%% and a~symmetric variant of the~Apollonios function~\cite{Santos:2013gra}.
%% The~double\nobreakdash-Gaussian function is taken as the~default model.
%% and the~other functions used to estimate the~systematic uncertainty.
%%
The  effective resolution depends on $m_{\Dp\Dm}$ and increases 
from $0.9\mevcc$ for $\decay{\Ppsi\mathrm{(3770)}}{\Dp\Dm}$
to $1.9\mevcc$ for $\decay{\Pchi_{\cquark2}\mathrm{(3930)}}{\Dp\Dm}$~signals
and is approximately $10\%$~larger for the~$\Dz\Dzb$~final state.
The~background in this region is found to be well described by a~second\nobreakdash-order polynomial function. 

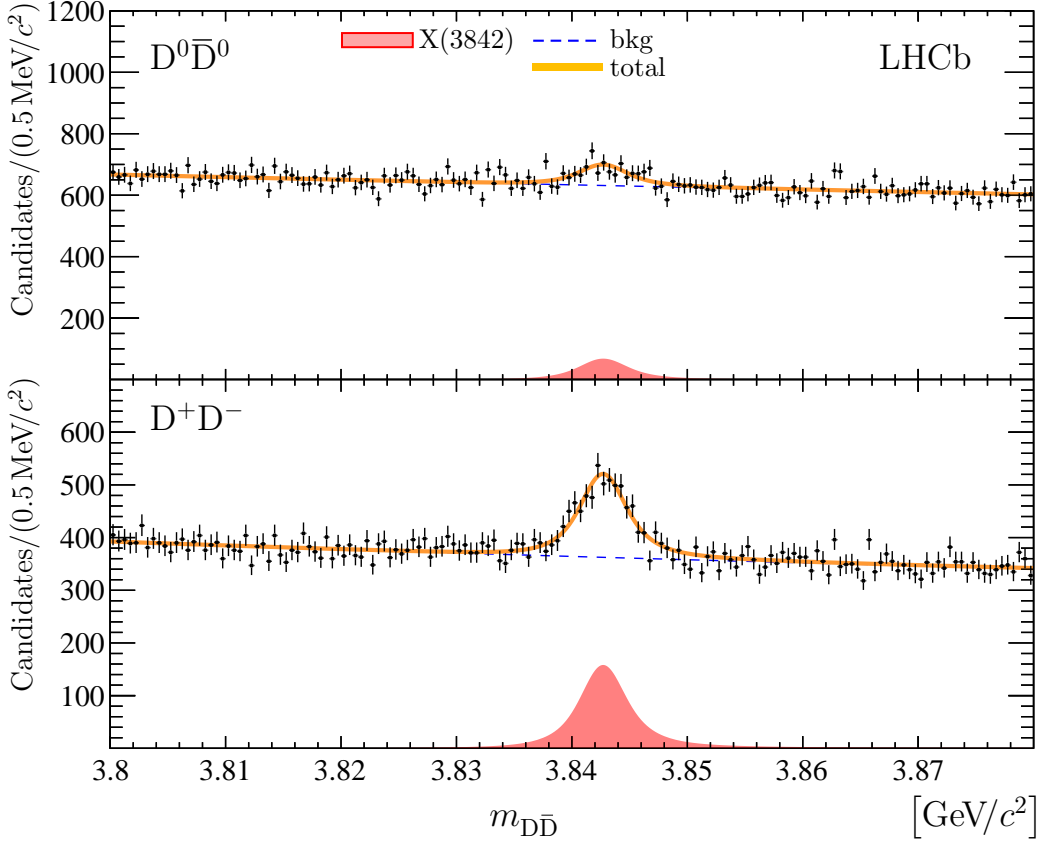
\begin{figure}[t]
  \setlength{\unitlength}{1mm}
  \centering
  \begin{picture}(150,120)
    %% 
    %% \graphpaper[5](-10,-10)(170,140)
    %% 
    \put(  0,  0){ 
      \includegraphics*[width=150mm,height=120mm,%
      %%]{NARROW_FIT_DM.pdf}
      ]{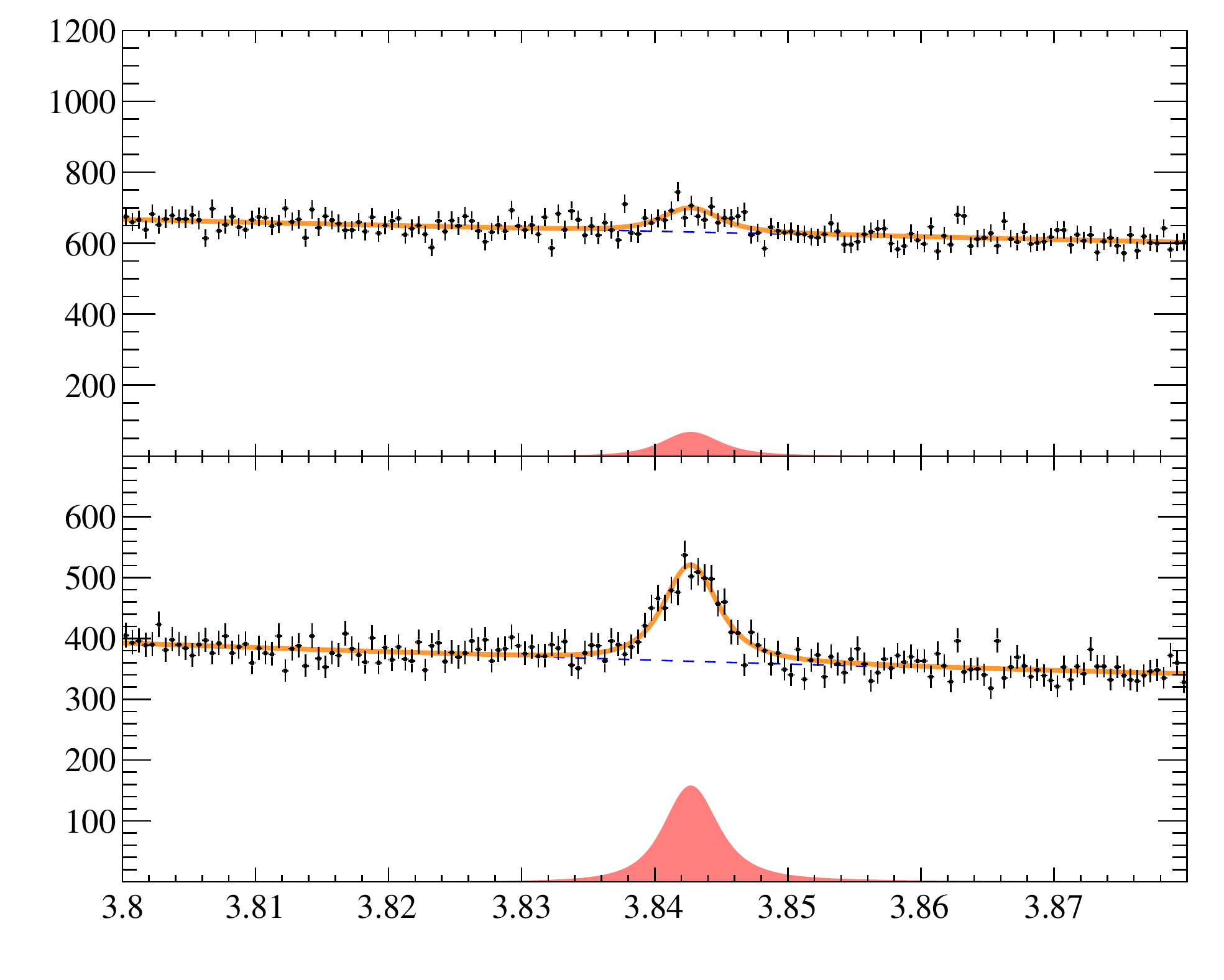}
    }
   \put(49,111) {\begin{tikzpicture}[x=1mm,y=1mm]\filldraw[fill=red!35!white,draw=red,thick]  (0,0) rectangle (10,2.0);\end{tikzpicture} }
    \put( 60,111) {\small$\mathrm{X(3842)}$}
    \put( 76,111) {\color[rgb]{0.00,0.00,1.00} {\hdashrule[0.5ex][x]{10mm}{0.3mm}{1.5mm 0.8mm} } } 
    \put( 76,107) {\color[rgb]{1.00,0.75,0.00} {\hdashrule[0.5ex][x]{10mm}{1.0mm}{1.0mm 0.0mm} } } 
    \put( 87,111) {\small$\mathrm{bkg}$}
    \put( 87,107) {\small{total} }
    \put( 22,108) {\large$\Dz\Dzb$ } 
    \put( 22, 57) {\large$\Dp\Dm$  }
    %%
    %%\put(110,107) {\small$\begin{array}{r}\mathrm{LHCb\ preliminary} \\ \sqs=7,8\&13\tev \\ 9\invfb \end{array}$} 
    \put(125,108){\large{LHCb}}
    %%\put(125,108){LHCb}
    %%\put(110,107) {\small$\begin{array}{r}\mathrm{LHCb} \\ \sqs=7,8\&13\tev \\ 9\invfb %%\end{array}$} 
    %%
    \put(  2, 73) {\begin{sideways}Candidates/(0.5\mevcc)\end{sideways}} 
    \put(  2, 20) {\begin{sideways}Candidates/(0.5\mevcc)\end{sideways}} 
    \put( 70,  1) {\large$m_{\D\Dbar}$}   \put(129, 1) {\large$\left[\!\gevcc\right]$} 
  \end{picture}
  \caption { \small
  Mass spectra of 
  (top)~$\Dz\Dzb$ and (bottom)~$\Dp\Dm$~candidates
in the~narrow \mbox{$3.80<m_{\D\Dbar}<3.88\gevcc$}~region. 
The~result of the~simultaneous fit described in the~text is superimposed.
  }
  \label{fig:dd_mass_narrow_fit}
\end{figure}

An~extended unbinned maximum\nobreakdash-likelihood fit is performed simultaneously to 
the~$\Dz\Dzb$ and $\Dp\Dm$~mass spectra. 
The~mass and the~natural width of the~$\mathrm{X(3842)}$~signals in the~$\Dz\Dzb$ and $\Dp\Dm$~final state are considered as common parameters in this fit whilst all other parameters are allowed to vary independently. 
All~parameters related to the~detector resolution are fixed to values found using simulation. 
The~result of the fit to 
the~data is shown in Fig.~\ref{fig:dd_mass_narrow_fit} and the~resulting parameters of 
interest are summarised in Table~\ref{tab:fits:dd_narrow}. 
The~statistical significance of the~$\mathrm{X(3842)}$~signal
is evaluated using Wilks' theorem~\cite{Wilks:1938dza} to be 
above $7\sigma$ for the $\Dz\Dzb$~decay mode
and 
above $21\sigma$ for the $\Dp\Dm$~decay mode. 

\begin{table}[h]        
  \centering            
  \caption{ \small
    Yields, mass and width of the~$\mathrm{X(3842)}$~state from 
    the~fit
    to $\D\Dbar$~mass spectra in 
    the~narrow \mbox{$3.80 < m_{\D\Dbar} <  3.88\gevcc$~region}.
    Uncertainties are statistical only.
  }\label{tab:fits:dd_narrow}
  \vspace*{3mm}
  \begin{tabular*}{0.75\textwidth}{@{\hspace{3mm}}l@{\extracolsep{\fill}}ccc@{\hspace{3mm}}}
    & $N_{\mathrm{X(3842)}}$
    & $m_{\mathrm{X(3842)}}$~$\left[\!\mevcc\right]$
    & $\Gamma_{\mathrm{X(3842)}}$~$\left[\!\mev\right]$
    \\[1mm]
    \hline 
    \\[-2mm]
    $\Dz\Dzb$
    %% & $\phantom{0}932\pm 165$ 
    & $\phantom{0}930\pm 170$
    & \multirow{2}{*}{$3842.71 \pm 0.16$}
    & \multirow{2}{*}{$\phantom{0}2.79\pm 0.51$}
    \\
    $\Dp\Dm$
    %% & $2072\pm 189$
    & $2070\pm190$
    &
    &
  \end{tabular*}
\end{table}

%\vspace*{1cm}
\subsection{\boldmath{Mass region~$3.80<m_{\D\Dbar}<4.20\gevcc$}\label{sec:chic}}

\begin{figure}[t]
  \setlength{\unitlength}{1mm}
  \centering
  \begin{picture}(150,120)
    %% 
    %% \graphpaper[5](-10,-10)(170,140)
    %%
    \put(  0,  0){ 
      \includegraphics*[width=150mm,height=120mm,%
      %%]{CHIC2_FIT_DM.pdf}
      ]{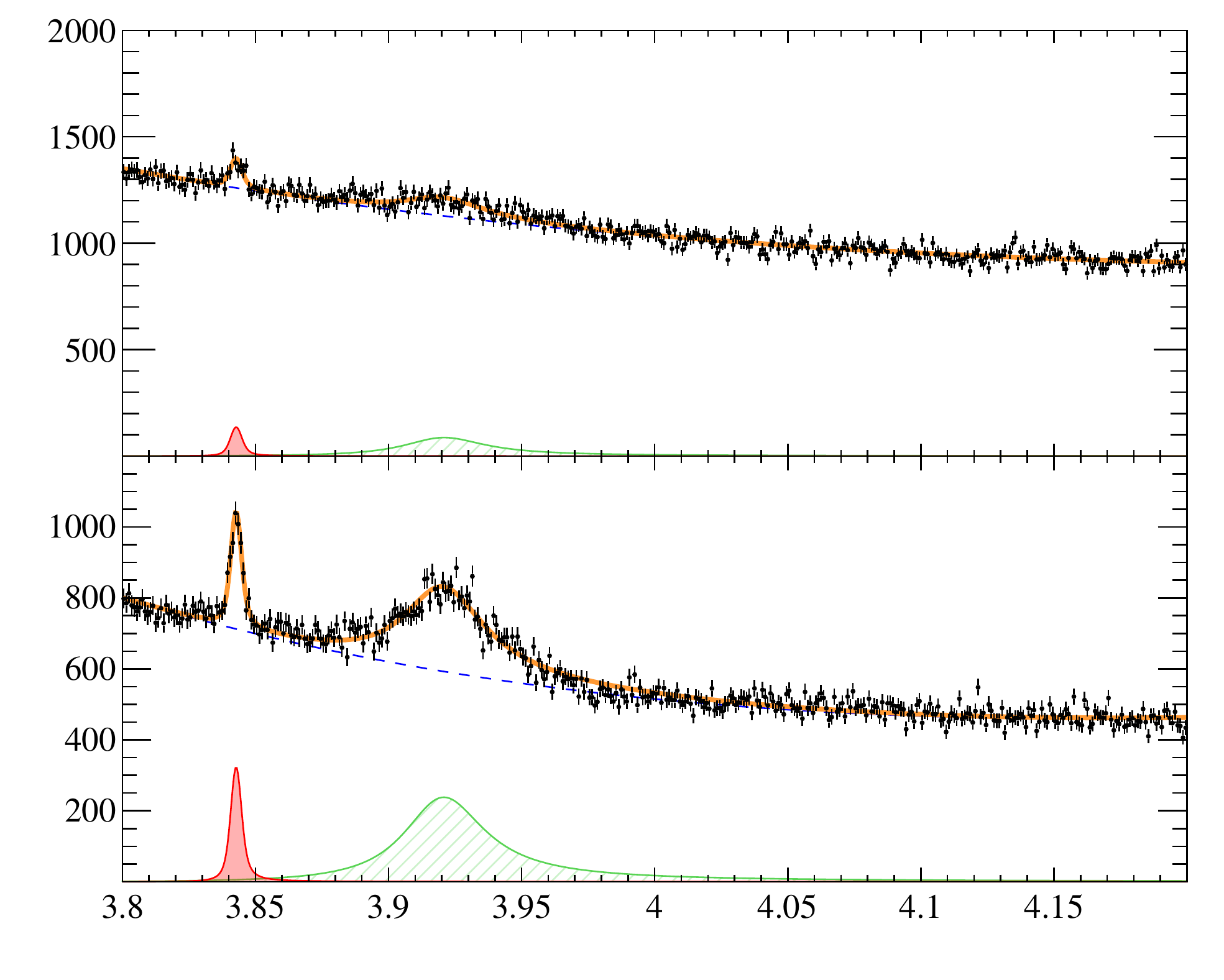}
    }
    %%
    %%\put( 46,111) {\color[rgb]{1.00,0.00,0.00}{\hdashrule[0.0ex][x]{10mm}{2.0mm}{1.0mm 0.0mm} } }
    \put(46,111) {\begin{tikzpicture}[x=1mm,y=1mm]\filldraw[fill=red!35!white,draw=red,thick]  (0,0) rectangle (10,2.0);\end{tikzpicture} }
    %%\put( 46,107) {\color[rgb]{1.00,0.00,1.00} {\hdashrule[0.5ex][x]{10mm}{0.5mm}{1.0mm 0.5mm 0.4mm 0.5mm} } } 
    \put( 57,111) {\small$\mathrm{X(3842)}$}
    \put( 57,107) {\small$\Pchi_{\cquark2}\mathrm{(3930)}$}
    \put( 76,111) {\color[rgb]{0.00,0.00,1.00} {\hdashrule[0.5ex][x]{10mm}{0.5mm}{1.0mm 1.0mm} } } 
    \put( 76,107) {\color[rgb]{1.00,0.75,0.00} {\hdashrule[0.5ex][x]{10mm}{1.0mm}{1.0mm 0.0mm} } } 
    \put( 87,111) {\small$\mathrm{bkg}$}
    \put( 87,107) {\small{total} }
    \put(46,107) {\begin{tikzpicture}[x=1mm,y=1mm]\draw[thick,green,pattern=north east lines, pattern color=green]  (0,0) rectangle (10,2.0);\end{tikzpicture} }
    \put( 32,108) {\large$\Dz\Dzb$ } 
    \put( 32, 57) {\large$\Dp\Dm$  }
    %%
    %%\put(110,107) {\small$\begin{array}{r}\mathrm{LHCb\ preliminary} \\ \sqs=7,8\&13\tev \\ 9\invfb \end{array}$} 
    %% \put(110,107) {\small$\begin{array}{r}\mathrm{LHCb} \\ \sqs=7,8\&13\tev \\ 9\invfb 
    %% \end{array}$}
    \put(125,108){\large{LHCb}}
    %%\put(105,108){LHCb preliminary}
    %%
    \put(  2, 73) {\begin{sideways}Candidates/(1\mevcc)\end{sideways}} 
    \put(  2, 21) {\begin{sideways}Candidates/(1\mevcc)\end{sideways}} 
    \put( 70,  1) {\large$m_{\D\Dbar}$}   \put(129, 1) {\large$\left[\!\gevcc\right]$} 
  \end{picture}
  \caption { \small
    Mass spectra of 
    (top)~$\Dz\Dzb$ and 
    (bottom)~$\Dp\Dm$~candidates
    in the~high\nobreakdash-mass \mbox{$3.80<m_{\D\Dbar}<4.20\gevcc$}~region. 
   The~result of the~simultaneous fit described in the~text is superimposed.
  }
  \label{fig:dd_mass_chic_fit}
\end{figure}

Two~signal components are used to describe the~$3.80<m_{\D\Dbar}<4.20\gevcc$~region:
the~$\mathrm{X(3842)}$~component, described earlier,
and a~component for the~$\Pchi_{\cquark2}\mathrm{(3930)}$~decay, modelled by the~convolution of a~relativistic D\nobreakdash-wave Breit\nobreakdash--Wigner function
with the~resolution model described above.
The~background in this mass region is modelled by an~exponential function 
multiplied by a~second\nobreakdash-order polynomial function.
The~total fit consists of the~sum of the~background
and the~$\mathrm{X(3842)}$
and $\Pchi_{\cquark2}\mathrm{(3930)}$~signals.
A~simultaneous extended binned maximum\nobreakdash-likelihood fit to 
the~$\Dz\Dzb$~and $\Dp\Dm$~mass spectra is performed with 
the~mass and natural width of the~$\mathrm{X(3842)}$~state
fixed to the results of the fit in 
the~narrow~\mbox{$3.80<m_{\D\Dbar}<3.88\gevcc$}~region.
%% Fixing the~mass and natural width of 
%% the~$\mathrm{X(3842)}$~state in the~fit
%% has no effect on the measurement of the~parameters 
%% of the~$\Pchi_{\cquark2}\mathrm{(3930)}$~resonance.
%%
The~mass and the natural width of the~$\Pchi_{\cquark2}\mathrm{(3930)}$~signals 
in the~$\Dz\Dzb$~and $\Dp\Dm$~final states 
and the~slope of the~background exponential function 
are common parameters and all other parameters 
are allowed to vary independently. 
The~result of the~fit of this model to the~data is shown in Fig.~\ref{fig:dd_mass_chic_fit} and the~resulting parameters of 
interest are summarised in Table~\ref{tab:fits:dd_chic}.
%%
%% If~the association of the~peak with the~$\Pchi_{\cquark2}\mathrm{(2P)}$~charmonium
%% state is incorrect and this state has spin~0, 
If~the~wide peak in Fig.~\ref{fig:dd_mass_chic_fit} 
is instead assumed to be spin\nobreakdash-0 then 
the~mass 
decreases by $0.12\mevcc$ while variations in the~width and the~uncertainties 
in the~mass and width are negligible.

\begin{table}[htb]
  \centering
  \caption{ \small
    Yields, mass and width of  the~$\Pchi_{\cquark2}\mathrm{(3920)}$~state
    from the~fit
    to $\D\Dbar$~mass spectra in 
    the~high\nobreakdash-mass \mbox{$3.88 < m_{\D\Dbar}<  4.20\gevcc$~region}.
    Uncertainties are statistical only.
  }\label{tab:fits:dd_chic}
  \vspace*{3mm}
  \begin{tabular*}{0.75\textwidth}{@{\hspace{3mm}}l@{\extracolsep{\fill}}ccc@{\hspace{3mm}}}
    & $N_{\Pchi_{\cquark2}\mathrm{(3930)}}~\left[10^3\right]$
    & $m_{\Pchi_{\cquark2}\mathrm{(3930)}}$~$\left[\!\mevcc\right]$
    & $\Gamma_{\Pchi_{\cquark2}\mathrm{(3930)}}$~$\left[\!\mev\right]$
    \\[1mm]
    \hline 
    \\[-2mm]
    $\Dz\Dzb$
    %% & $\phantom{0}4718\pm526\phantom{0}$
    %% & $\phantom{0}4700\pm500\phantom{0}$    %% PDG rounding   
    & $\phantom{0}4.7\pm0.5$                   %% PDG rounding  
    & \multirow{2}{*}{$3921.90 \pm 0.55$}
    & \multirow{2}{*}{$36.64   \pm 1.88$}
    \\
    $\Dp\Dm$
    %% & $12980\pm559\phantom{0}$
    %% & $1300\pm600\phantom{0}$               %% PDG rounding 
    & $13.0\pm0.6$                             %% PDG rounding 
    & 
    &
  \end{tabular*}
\end{table}

%%

%\vspace*{1cm}
\subsection{\boldmath{Mass region~$m_{\D\Dbar}<3.88\gevcc$}}\label{sec:psi}

To fit the~$\D\Dbar$~mass spectra in the~near\nobreakdash-threshold region, \mbox{$m_{\D\Dbar}<3.88\gevcc$}, 
components for the~$\mathrm{X(3842)}$ and $\Ppsi\mathrm{(3770)}$~decays to $\D\Dbar$~signals
and the background are included.
In~the~case of the~$\Dz\Dzb$~mass spectrum, 
an~additional contribution from 
$\decay{\Pchi_{\cquark1}\mathrm{(3872)}}{\D{}^{\ast0}\Dzb}$~decays 
followed by $\decay{\D{}^{\ast0}}{\Dz\piz}$~or $\decay{\D{}^{\ast0}}{\Dz\g}$~is required.
The~$\decay{\Ppsi\mathrm{(3770)}}{\D\Dbar}$~component is modelled as a~relativistic 
\mbox{multi}\nobreakdash-channel P\nobreakdash-wave 
\mbox{Breit}\nobreakdash--Wigner function~\cite{Anashin:2011kq,Ablikim:2007gd}, accounting for decays into $\Dz\Dzb$, 
$\Dp\Dm$ and non\nobreakdash-$\D\Dbar$~final states~\cite{PDG2018},
convolved with a~double\nobreakdash-Gaussian resolution model. 
The~background is modelled as a~product of a~scaled two\nobreakdash-body phase\nobreakdash-space function and a~second\nobreakdash-order polynomial function.
The~shape of the~feed\nobreakdash-down contribution from $\Pchi_{\cquark1}\mathrm{(3872)}$~decays is described using simulated 
two\nobreakdash-body $\decay{\Pchi_{\cquark1}\mathrm{(3872)}}{\D{}^{\ast0}\Dzb}$
and three\nobreakdash-body $\decay{\Pchi_{\cquark1}\mathrm{(3872)}}{\Dz\Dzb\piz}$~decays.
The~latter corresponds to off\nobreakdash-shell decays of the~intermediate $\D{}^{\ast0}$~mesons~\cite{Braaten:2007dw,Stapleton:2009ey}.
The~simulation of $\decay{\Pchi_{\cquark1}\mathrm{(3872)}}{\D{}^{\ast0}\Dzb}$~decays
%% In~the~simulation the~$\D{}^{\ast0}$~mesons
assumes that  the~$\D{}^{\ast0}$~mesons are unpolarised
and the~three-body decay dynamics are not included.
The~contributions from the~two\nobreakdash-body and three\nobreakdash-body 
decays of the~$\Pchi_{\cquark1}\mathrm{(3872)}$~state
are allowed to vary independently in the~fit. 

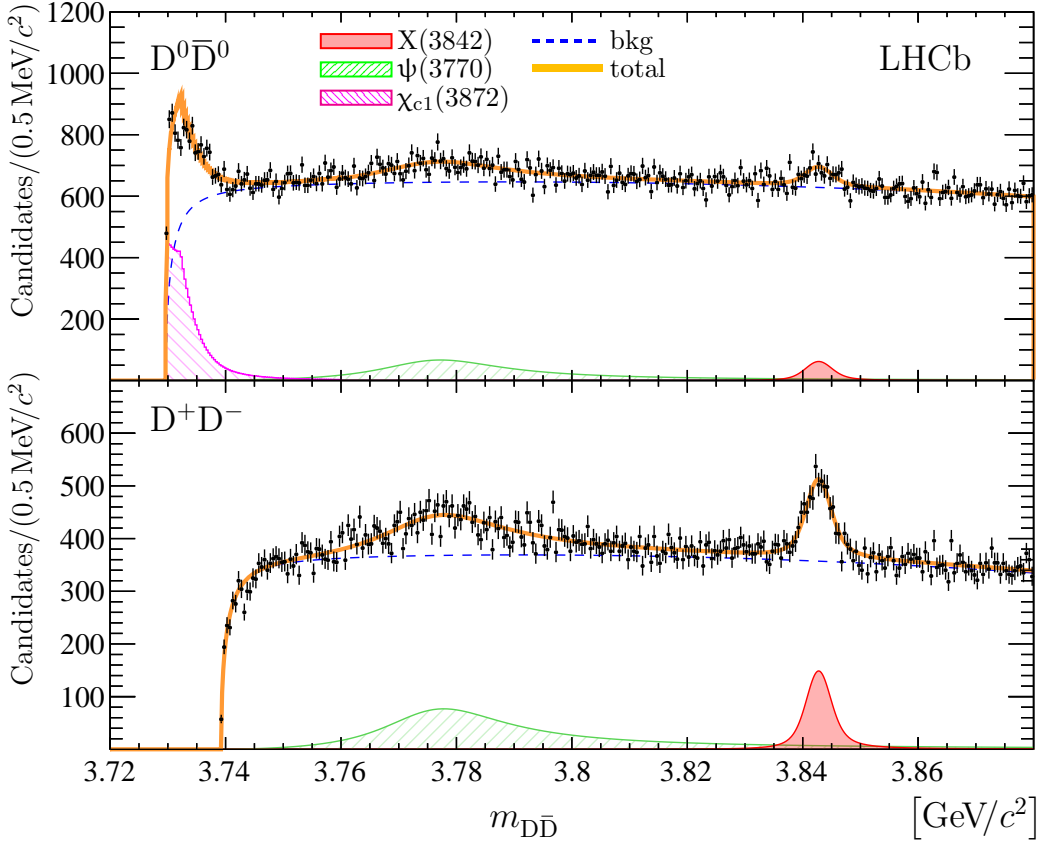
\begin{figure}[t]
  \setlength{\unitlength}{1mm}
  \centering
  \begin{picture}(150,120)
    %% 
    %%\graphpaper[5](-10,-10)(170,140)
    %% 
    \put(  0,  0){ 
      \includegraphics*[width=150mm,height=120mm,%
      %% ]{PSI_FIT_DM.pdf}
      ]{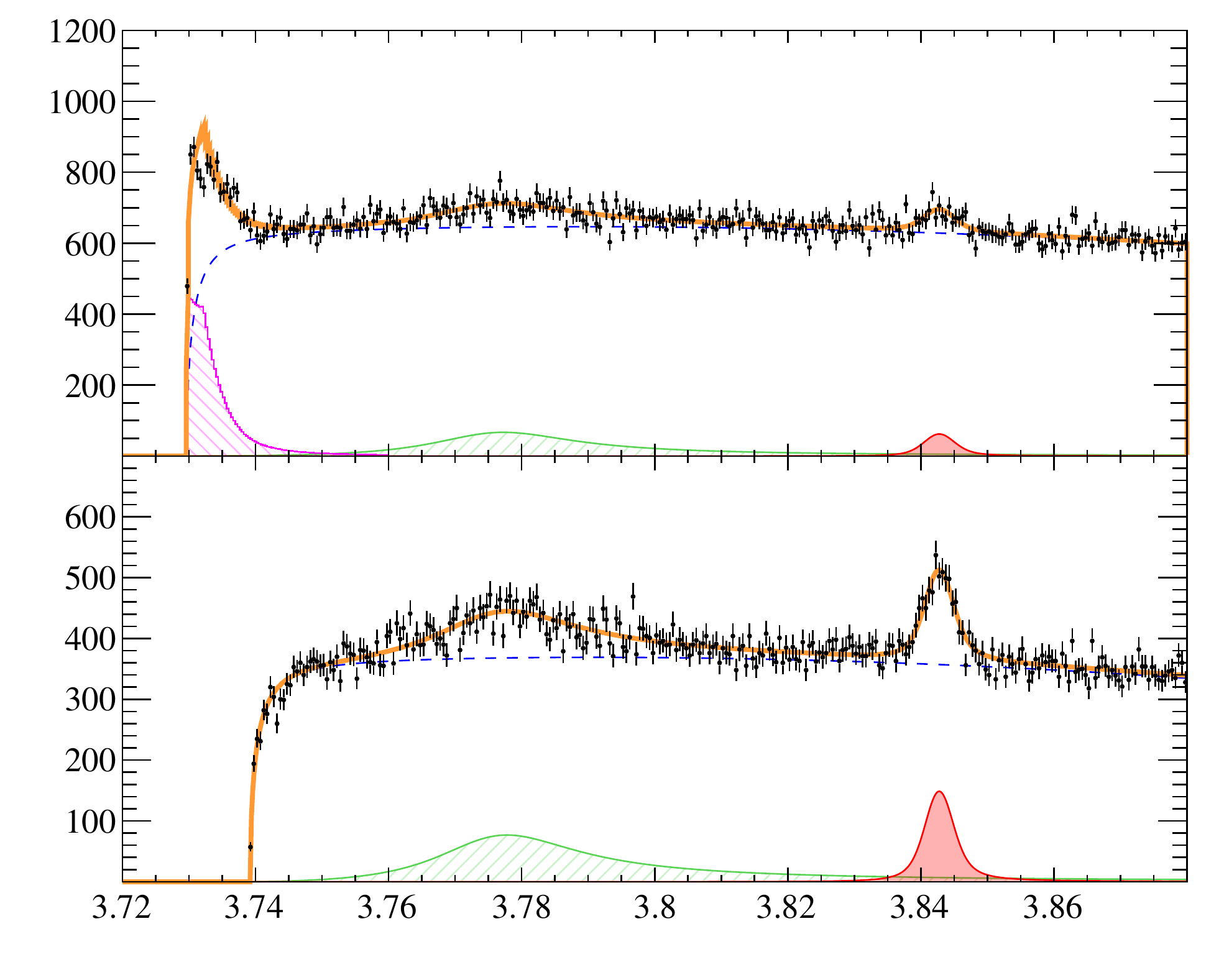}
    }
    %%
    %%\put( 46,111) {\color[rgb]{1.00,0.00,0.00} {\hdashrule[0.0ex][x]{10mm}{2.0mm}{1.0mm 0.0mm} } }
    %%
    \put(46,111) {\begin{tikzpicture}[x=1mm,y=1mm]\filldraw[fill=red!35!white,draw=red,thick]  (0,0) rectangle (10,2.0);\end{tikzpicture} }
    \put( 57,111) {\small$\mathrm{X(3842)}$}
    \put( 57,107) {\small$\Ppsi\mathrm{(3770)}$}
    \put( 57,103) {\small$\Pchi_{\cquark1}\mathrm{(3872)}$}
    \put( 76,111) {\color[rgb]{0.00,0.00,1.00} {\hdashrule[0.5ex][x]{10mm}{0.5mm}{1.0mm 1.0mm} } } 
    \put( 76,107) {\color[rgb]{1.00,0.75,0.00} {\hdashrule[0.5ex][x]{10mm}{1.0mm}{1.0mm 0.0mm} } } 
    \put( 87,111) {\small$\mathrm{bkg}$}
    \put( 87,107) {\small{total} }
    \put(46,107) {\begin{tikzpicture}[x=1mm,y=1mm]\draw[thick,green,pattern=north east lines, pattern color=green]  (0,0) rectangle (10,2.0);\end{tikzpicture} }
    \put(46,103) {\begin{tikzpicture}[x=1mm,y=1mm]\draw[thick,magenta,pattern=north west lines, pattern color=magenta]  (0,0) rectangle (10,2.0);\end{tikzpicture} }
    \put( 22,108) {\large$\Dz\Dzb$ } 
    \put( 22, 57) {\large$\Dp\Dm$  }
    %%
    %%\put(110,107) {\small$\begin{array}{r}\mathrm{LHCb\ preliminary} \\ \sqs=7,8\&13\tev \\ 9\invfb \end{array}$} 
    %%\put(110,107) {\small$\begin{array}{r}\mathrm{LHCb} \\ \sqs=7,8\&13\tev \\ 9\invfb %%\end{array}$} 
    \put(125,108){\large{LHCb}}
    %%\put(105,108){LHCb preliminary}
    %%
    \put(  2, 73) {\begin{sideways}Candidates/(0.5\mevcc)\end{sideways}} 
    \put(  2, 21) {\begin{sideways}Candidates/(0.5\mevcc)\end{sideways}} 
    \put( 70,  1) {\large$m_{\D\Dbar}$}   \put(129, 1) {\large$\left[\!\gevcc\right]$} 
  \end{picture}
  \caption { \small
    Mass spectra of 
       (top)~$\Dz\Dzb$ and 
    (bottom)~$\Dp\Dm$~candidates
    in the~near\nobreakdash-threshold 
    \mbox{$m_{\D\Dbar}<3.88\gevcc$}~region.
     The~result of the~simultaneous fit described in the~text is superimposed.
  }
  \label{fig:dd_mass_psi_fit}
\end{figure}

A~simultaneous binned extended maximum\nobreakdash-likelihood fit to the~$\Dz\Dzb$~and 
$\Dp\Dm$~mass spectra is performed. In~this fit, the~mass and width of 
the~$\mathrm{X(3842)}$~signal are fixed  from 
the~results of the~unbinned fit in the~narrow~\mbox{$3.80<m(\D\Dbar)<3.88\gevcc$}~region, 
the~mass of the~$\Ppsi\mathrm{(3770)}$~state is allowed to vary, while the~natural width of the~$\Ppsi\mathrm{(3770)}$~state is Gaussian\nobreakdash-constrained to the~known value  of $\Gamma_{\Ppsi\mathrm{(3770)}}=27.2\pm1.0\mev$~\cite{PDG2018}.
The~mass of the~$\Ppsi\mathrm{(3770)}$~state
and the~scale factor for the~background two\nobreakdash-body phase space function 
are common parameters and all other parameters are allowed to vary independently.
The~result of the~fit 
to the~$\Dz\Dzb$ and $\Dp\Dm$~mass spectra is shown 
in Fig.~\ref{fig:dd_mass_psi_fit} and the~resulting parameters of 
interest are summarised in Table~\ref{tab:fits:dd_psi}.
The~fit~quality in the~region~\mbox{$m_{\Dz\Dzb}<3.74\gevcc$} is poor, 
possibly due to large effects of the~neglected dynamics in $\decay{\Pchi_{\cquark1}\mathrm{(3872)}}{\Dz\Dzb\mathrm{X}}$~decays. 
%%
%%  It is found that large variations of the shape of
%% the~$\Pchi_{\cquark1}\mathrm{(3872)}$~component 
%% do not affect the~measured mass of the~$\Ppsi\mathrm{(3770)}$~state.
However, it is found that the~exact description 
of  the~$\Pchi_{\cquark1}\mathrm{(3872)}$~contribution 
does not affect the~measurement of the~mass of the~$\Ppsi\mathrm{(3770)}$~state.

\begin{table}[tb]
  \centering
  \caption{ \small
      Yields and mass of the~$\Ppsi\mathrm{(3770)}$~state from
        the~fit
    to $\D\Dbar$~mass spectra in 
    the~near\nobreakdash-threshold \mbox{$m_{\D\Dbar} <  3.88\gevcc$~region}.
    Uncertainties are statistical only.
  }\label{tab:fits:dd_psi}
  \vspace*{3mm}
  \begin{tabular*}{0.65\textwidth}{@{\hspace{3mm}}l@{\extracolsep{\fill}}cc@{\hspace{3mm}}}
    & $N_{\Ppsi\mathrm{(3770)}}~\left[10^3\right]$
    & $m_{\Ppsi\mathrm{(3770)}}$~$\left[\!\mevcc\right]$
    \\[1mm]
    \hline 
    \\[-2mm]
    $\Dz\Dzb$
    %% & $5116\pm547$ 
    & $5.1\pm0.5$
    & \multirow{2}{*}{$3778.13 \pm 0.70$}
    \\
    $\Dp\Dm$
    %% & $5719\pm436$
    & $5.7\pm0.4$          %% PDG rounding
    & 
  \end{tabular*}
\end{table}

%%\clearpage
\section{Systematic uncertainties}
\label{sec:systematics}
In~the proximity of the~$\D\Dbar$ mass threshold 
most potential systematic uncertainties for the~mass and natural width measurements 
become negligible when $\D$~mass constraints are applied. 
The~main systematic uncertainties for 
the~measured~$\mathrm{X(3842)}$, $\Pchi_{\cquark2}\mathrm{(3930)}$
and $\Ppsi\mathrm{(3770)}$~resonance parameters are related
to the~signal and background parameterisation,
the~momentum\nobreakdash-scale calibration and the~uncertainty in 
the~known $\Dz$~and  $\Dp$~masses~\cite{PDG2018}. 
These are described below and summarised 
in Table~\ref{tab:systematic}.

\begin{table}[t]
  \centering
  \caption{\small 
    Summary of systematic uncertainties for  
    the~measured masses\,($\upsigma_m$) 
    and width\,($\upsigma_\Gamma$) of~the~$\mathrm{X(3842)}$, $\chi_{\cquark2}\mathrm{(3930)}$
    and $\Ppsi\mathrm{(3770)}$~states.
    Uncertainties for the mass\,(width) smaller than $10\kevcc\,(10\kev)$~are not  shown.
  }\label{tab:systematic}
  \vspace*{3mm}
  \begin{tabular*}{0.98\textwidth}{@{\hspace{1mm}}l@{\extracolsep{\fill}}ccccc@{\hspace{1mm}}}
  \multirow{3}{*}{~~~Source}
    & \multicolumn{2}{c}{$\mathrm{X(3842)}$}
    & \multicolumn{2}{c}{$\Pchi_{\cquark2}\mathrm{(3930)}$}
    & $\Ppsi\mathrm{(3770)}$
    \\
    &  $\upsigma_m$ 
    &  $\upsigma_\Gamma$ 
    &  $\upsigma_m$
    &  $\upsigma_\Gamma$
    &  $\upsigma_m$
    \\
    & $\left[\!\mevcc\right]$
    & $\left[\!\mev\right]$
    & $\left[\!\mevcc\right]$
    & $\left[\!\mev\right]$
    & $\left[\!\mevcc\right]$
    \\[1mm]
    \hline 
    \\[-2mm]
    Signal model     
    &  $0.02$
    &  $0.02$
    &  $0.01$
    &  $0.15$
    &  $0.62$
    \\
    Resolution
    &  
    &  $0.31$
    & 
    &  $0.20$
    &  
    \\
    Background model       
    &  
    &  $0.13$ 
    &  $0.15$
    &  $0.81$
    &  $0.03$
    \\
    Momentum scale 
    &   $0.07$
    &   --- 
    &   $0.05$
    &   --- 
    &   
    \\
    $\D$\nobreakdash-meson masses 
    &  $0.10$
    &   ---
    &  $0.10$
    &   ---
    &  $0.10$
   \\[1mm]
    \hline 
    \\[-3mm]
    %%Total
    Sum in   quadrature
    &  $0.12$
    &  $0.35$
    &  $0.19$
    &  $0.85$
    &  $0.63$
\end{tabular*}
\end{table}

To~evaluate the~systematic uncertainty related to the~parameterisation of the~signal shape,
the~parameters of the~relativistic Breit\nobreakdash--Wigner functions are varied.
In~particular, the~meson radius, entering the~Blatt\nobreakdash--Weisskopf 
centrifugal factor
with the~default value of $3.5\gev^{-1}$, is varied between 
$1.5\gev^{-1}$ and $5\gev^{-1}$.
In the case of~the~$\mathrm{X(3842)}$~state, where the~quantum numbers are unknown,
the~orbital momentum is varied between zero and four.
For~the~$\mathrm{X(3842)}$ and $\Pchi_{\cquark2}\mathrm{(3930)}$~states,
alternative signal descriptions with multi\nobreakdash-channel
relativistic Breit\nobreakdash--Wigner functions with 
$\Dz\Dzb$~and $\Dp\Dm$  and radiative non\nobreakdash-$\D\Dbar$~decays are used.
For~the~$\Ppsi\mathrm{(3770)}$~signal, the~parameters of 
the~multi\nobreakdash-channel relativistic P\nobreakdash-wave 
\mbox{Breit}\nobreakdash--\mbox{Wigner} function, namely the ratio 
of branching fractions to $\Dz\Dzb$~and $\Dp\Dm$~final states, 
and the~branching fraction for non\nobreakdash-$\D\Dbar$, are  
varied within their~known uncertainties~\cite{PDG2018}. 

The~determination of the natural width of 
the~$\mathrm{X(3842)}$  and $\Pchi_{\cquark2}\mathrm{(3930)}$~states 
relies on accurate modelling  of the~detector resolution.
Comparing data and simulation for decay modes with 
low energy release such as 
the~$\decay{\Pchi_{\cquark1}}{\jpsi\mumu}$~decay,
agreement 
at the~$10\%$ level is found~\cite{LHCb-PAPER-2017-036}. 
Even~better agreement is found for $\bquark$\nobreakdash-hadron decays 
to pairs of open charm hadrons such as 
$\decay{\Bz}{\Ds\Dm}$,
$\decay{\Lb}{\Lc\Dsm}$ and
$\decay{\Lb}{\Lc\Dm}$~\cite{LHCb-PAPER-2014-002}, 
where 
the~energy release is larger. Hence, to estimate the~corresponding uncertainty 
the~resolution scale is varied by $10\%$ and the~fit is repeated.
Alternative resolution models, such as 
a~symmetric double-sided 
Crystal Ball function~\cite{Skwarnicki:1986xj,LHCB-Paper-2011-013}
and 
a~symmetric variant of the~Apollonios function~\cite{Santos:2013gra}
are used to estimate the~uncertainty associated 
with this choice.

%% Three~resolution models are  found to describe simulated data well:
%% a~double\nobreakdash-Gaussian function, 
%% a~symmetric double\nobreakdash-sided Crystal Ball 
%% function~\cite{Skwarnicki:1986xj,LHCB-Paper-2011-013} 
%% and a~symmetric variant of the~Apollonios function~\cite{Santos:2013gra}.
%% The~double\nobreakdash-Gaussian function is taken as the~default model.
%% and the~other functions used to estimate the~systematic uncertainty.
%

%% {\color{red}{
    The~uncertainty in the~knowledge of the~width of the~$\Ppsi\mathrm{(3770)}$~resonance~\cite{PDG2018}
    is propagated by applying a~Gaussian constraint in the~fit,
    and it is therefore a~part of the~statistical uncertainty for
    the~measured mass of the~$\Ppsi\mathrm{(3770)}$~state.
    The~effect of fixing the~parameters of the~$\mathrm{X(3842)}$ state in the~fits
    in the~\mbox{$m_{\D\Dbar}<3.88\gevcc$} and \mbox{$m_{\D\Dbar}>3.8\gevcc$}~regions
    on the~parameters of the~$\Pchi_{\cquark2}\mathrm{(3930)}$ and
    $\Ppsi\mathrm{(3770)}$~states is found to be negligible.
%% }}
The~effect of the poorly known shape for
the~$\decay{\Pchi_{\cquark1}\mathrm{(3872)}}{\Dz\Dzb\mathrm{X}}$~component
has no visible effect on
the~determination of the~mass of the~$\Ppsi\mathrm{(3770)}$~state.

The~impact of the~choice of the~background model is estimated by 
changing the~order of 
the~polynomial functions from second to fourth order
and, for fits in the~\mbox{$3.80<m_{\D\Dbar}<3.88\gevcc$}
and~\mbox{$m_{\D\Dbar}<3.88\gevcc$}~regions,
by including an~exponential factor to the background model.
For~the fit in the~\mbox{$3.80<m_{\D\Dbar}<3.88\gevcc$}~region,
the~contributions from the~long tails of 
the~wide $\Ppsi\mathrm{(3770)}$ and $\Pchi_{\cquark2}\mathrm{(3930)}$~resonances
are accounted for.

The Particle Data Group\,(PDG)~\cite{PDG2018} reports various heavy or exotic 
charmonium candidates that decay to $\D\Dbar$, 
$\D{}^{\ast}\Dbar$ and $\D{}^{\ast}\Dbar{}^{\ast}$~final states. 
Typically, these states are relatively broad and consequently 
they will only be visible as a~distortion of the~background shape. 
To~study the~impact of these charmonium states on the~measurements made here,
the~decays $\decay{\mathrm{Z}_{\cquark}\mathrm{(3900)}}{\Dz{}\D{}^{\ast-}}$, 
$\decay{\mathrm{X(4020)}}{\D{}^{\ast}\Dbar{}^{\ast}}$,
$\decay{\Pchi_{\cquark0}\mathrm{(3860)}}{\D\Dbar}$,
and decays of 
$\Ppsi\mathrm{(4040)}$,
$\Ppsi\mathrm{(4160)}$,
$\Ppsi\mathrm{(4415)}$
to $\D\Dbar$,
$\D{}^{\ast}\Dbar$
and 
$\D{}^{\ast}\Dbar{}^{\ast}$~final states~\cite{PDG2018}
are simulated and 
%% their contributions to $\Dz\Dzb$~and $\Dp\Dm$~mass spectra
%% are added 
individually added as fit components in turn.
For these studies, 
the~measurements 
of the~relative direct\,($\D\Dbar$) and 
feed-down\,($\D{}^{\ast}\Dbar$ and 
$\D{}^{\ast}\Dbar{}^{\ast}$) contributions~\cite{PDG2018}
provide important constraints.
%%
%% In~these studies the~information available from the~direct 
%% and feed\nobreakdash-down contributions %~\cite{PDG2018} 
%% to the~modes considered here provides important constraints.
Fits including decays of 
the~$\Pchi_{\cquark0}\mathrm{(3860)}$, $\Ppsi\mathrm{(4040)}$ 
or $\Ppsi\mathrm{(4160)}$~states are found to modify
the~background component and cause a~maximum of $0.15\mevcc$ bias on the~mass 
and a~maximum of $0.5\mev$~bias on the~natural width of the~$\Pchi_{\cquark2}\mathrm{(3930)}$~state.
These are accounted for as uncertainties due to 
the~background description. Contributions from other charmonium or 
charmonium\nobreakdash-like states have no effect in the~determination of the~parameters of the~$\mathrm{X(3842)}$, $\Pchi_{\cquark2}\mathrm{(3930)}$ and $\Ppsi\mathrm{(3770)}$~states.

An important experimental uncertainty for the~mass measurements is the~knowledge of the~momentum scale. This~is minimised by the~application of the~$\D$\nobreakdash-mass constraints. 
The~residual uncertainty from this source is evaluated by adjusting the~momentum scale by 
the~$3 \times 10^{-4}$ uncertainty on the calibration procedure and repeating the~mass fit.
A~further uncertainty of $0.1\mevcc$ arises from the knowledge of the~$\Dz$ and $\Dp$~masses~\cite{PDG2018}.

%%\clearpage

\vspace{1cm}
\section{Production mechanism}\label{sec:production}

The~selection criteria used in this analysis significantly suppress a potential contribution from weak decays of long\nobreakdash-lived beauty hadrons. 
To~probe the residual contribution from $\bquark$-hadron decays,
the~sample of $\D\Dbar$~pairs is split into two subsamples 
according to the~value of the~$t_{z}$~variable~\cite{LHCb-PAPER-2011-003}
\begin{equation*}
  t_{z} \equiv \dfrac {z_{\D\Dbar} - z_{\mathrm{PV}} }  { p_z} m_{\D\Dbar}\,,
\end{equation*}
%%\mbox{$m_{\D\Dbar}(z_{\D\Dbar} - z_{\mathrm{PV}})/p_z$},
where $z_{\D\Dbar}$ and $z_{\mathrm{PV}}$~are
the~positions along the~$z$\nobreakdash-axis\,(the beam direction) 
of the~reconstructed $\D\Dbar$~vertex and of the~primary vertex,
and  $p_z$~is the~measured $\D\Dbar$~momentum in the~$z$~direction.
Promptly produced charmonia are characterised by a~nearly symmetric and narrow distribution around~\mbox{$t_z=0$},
whilst almost all $\D\Dbar$~pairs being produced in the~weak decays of long\nobreakdash-lived beauty hadrons have $t_z>0$.
Comparison of the~observed yields of the~$\mathrm{X(3842)}$, 
$\Pchi_{\cquark2}\mathrm{(3930)}$ and 
$\Ppsi\mathrm{(3770)}$~signals for~$t_z<0$ and 
$t_z>0$~subsamples shows no sizeable contributions from decays 
of $\bquark$~hadrons 
to the~$\mathrm{X(3842)}$ and $\Pchi_{\cquark2}\mathrm{(3930)}$~signals, 
while a~contribution of~$\sim35\%$ to the~observed yield of 
the~$\decay{\Ppsi\mathrm{(3770)}}{\D\Dbar}$~decays is found.

Reference~\cite{Barnes:2005pb} suggests 
the decay~$\decay{\Pchi_{\cquark2}\mathrm{(2^3P_2)}}{\Ppsi_3\mathrm{(1^3D_3)\g}}$
as a~possible production mechanism for the~$\Ppsi_3\mathrm{(1^3D_3)}$~state. 
The~hypothesis is tested as follows.
Identifying the~$\Pchi_{\cquark2}\mathrm{(3930)}$ 
as~$\Pchi_{\cquark2}\mathrm{(2^3P_2)}$
and  $\mathrm{X(3842)}$ as 
$\Ppsi_3\mathrm{(1^3D_3)}$ and taking
$\Gamma\left(\decay{\Pchi_{\cquark2}\mathrm{(2^3P_2)}}{\Ppsi_3\mathrm{(1^3D_3)\g}}\right)$ 
to be $100\kev$~\cite{Barnes:2005pb},
from the present measurement of the~$\Pchi_{\cquark2}\mathrm{(3930)}$~state width 
and  the~observed yields of 
$\decay{\Pchi_{\cquark2}\mathrm{(3930)}}{\D\Dbar}$~decays, at most $5\%$~of the~observed 
$\decay{\mathrm{X(3842)}}{\D\Dbar}$~decays can originate from the~decays of the~$\Pchi_{\cquark2}\mathrm{(3930)}$~state. 
This~suggests, assuming the~$\Ppsi_3\mathrm{(1^3D_3)}$ assignment is correct, that either $\Gamma\left(\decay{\Pchi_{\cquark2}\mathrm{(2^3P_2)}}{\Ppsi_3\mathrm{(1^3D_3)\g}}\right)$ is 
significantly larger than expected or that a~large fraction of 
the~$\mathrm{X(3842)}$~signal is produced via a~different production mechanism.

%%\clearpage

\section{Results and discussion}\label{sec:results}
Using the~LHCb dataset collected between 2011 and 2018, 
near-threshold~$\D\Dbar$~mass spectra are studied and 
a~new narrow charmonium state, the $\mathrm{X(3842)}$, is observed in the~decay modes $\decay{\mathrm{X(3842)}}{\Dz\Dzb}$ and 
$\decay{\mathrm{X(3842)}}{\Dp\Dm}$ with very high statistical significance. 
The~mass and the~natural width of this state are measured to be 
\begin{eqnarray*}
  m_{\mathrm{X(3842)}}  & = & 3842.71           \pm 0.16 \pm 0.12 \mevcc\,, \\
  \Gamma_{\mathrm{X(3842)}} & = & \phantom{000}2.79 \pm 0.51 \pm 0.35 \mev\,,
\end{eqnarray*}
where the~first uncertainty is statistical and the~second is systematic.
The~narrow natural width and measured value of the mass suggests the~interpretation of the~$\mathrm{X(3842)}$~state as the~$\Ppsi_3\mathrm{\left(1^3D_3\right)}$~charmonium state with $\mathrm{J^{PC}=3^{--}}$.

In addition, prompt hadroproduction of the~$\Pchi_{\cquark2}\mathrm{(3930)}$~state is observed 
for the~first time, and the~parameters of this~state are measured to be 
\begin{eqnarray*}
  %% m_{\Pchi_{\cquark2}\mathrm{(3930)}}  & = & 3921.90 \pm 0.55 \pm 0.19 \mevcc\,, \\ 
  %% \Gamma_{\Pchi_{\cquark2}\mathrm{(3930)}} & = &\phantom{00}36.64 \pm 1.88 \pm 0.85 \mev\,.
  m_{\Pchi_{\cquark2}\mathrm{(3930)}}  & = & 3921.9 \pm 0.6 \pm 0.2 \mevcc\,, \\ 
  \Gamma_{\Pchi_{\cquark2}\mathrm{(3930)}} & = &\phantom{00}36.6 \pm 1.9 \pm 0.9 \mev\,.
\end{eqnarray*}
These values are considerably more precise than previous measurements made at~$\mathrm{e}^+\mathrm{e}^-$~machines, as can be seen from Table~\ref{tab:results:chic}. The~mass measured in this analysis  is $2 \sigma$ lower than the~current world average whilst the~natural width is $2 \sigma$ higher. It~is interesting to note that the~measured value of the~mass is roughly midway between the~masses quoted in Ref.~\cite{PDG2018} for this state and and for the~$\mathrm{X(3915)}$~meson, which is only known to decay to 
the~$\jpsi\Pomega$~final state~\cite{Lees:2012xs,delAmoSanchez:2010jr,Uehara:2009tx,Abe:2004zs,Adachi:2008sua}. 
Further studies are needed to understand if there are two distinct charmonium states in this region or only one as suggested in Ref.~\cite{Zhou:2015uva}.

\begin{table}[tb]
  \centering
  \caption{ \small
    Summary of mass and width measurements for the~$\Pchi_{\cquark2}\mathrm{(3930)}$~state.
  }\label{tab:results:chic}
  \vspace*{3mm}
  \begin{tabular*}{0.80\textwidth}{@{\hspace{5mm}}l@{\extracolsep{\fill}}cll@{\hspace{5mm}}}
    & 
    & $~~m_{\Pchi_{\cquark2}\mathrm{(3930)}}~\left[\!\mevcc\right]$
    & $~~\Gamma_{\Pchi_{\cquark2}\mathrm{(3930)}}~\left[\!\mev\right]$
    \\[1mm]
    \hline 
    \\[-2mm]
    Belle & \cite{Uehara:2005qd} & $3929\phantom{.0} \pm 5\phantom{.0} \pm 2\phantom{.0} $  & $29\phantom{.0}\pm10\phantom{.}\pm2\phantom{.0}$ 
    \\
    BaBar & \cite{Aubert:2010ab} 
    & $3926.7 \pm 2.7 \pm 1.1 $  
    & $21.3\pm6.8\pm3.6$
    \\
    %%PDG average & \cite{PDG2018}         
    %% & $3927.2\phantom{0} \pm 2.6\phantom{0} \phantom{\pm 0.00}$   
    %% & $24\phantom{.00} \pm 6\phantom{.00}\phantom{\pm0.00} $
    %% \\
    This analysis & 
    %% &  $3921.90 \pm 0.55 \pm 0.19$
    %% &  $36.64 \pm 1.88 \pm 0.85$ 
    &  $3921.9 \pm 0.6 \pm 0.2$
    &  $36.6 \pm 1.9 \pm 0.9$ 
  \end{tabular*}
\end{table}

Finally, prompt hadroproduction of the $\Ppsi\mathrm{(3770)}$~state is observed for the~first time, and the~mass of this state is measured to be 
\begin{equation*}
  %% m_{\Ppsi\mathrm{(3770)}}  = 3778.13 \pm 0.70 \pm 0.63 \mevcc\,. 
   m_{\Ppsi\mathrm{(3770)}}  = 3778.1 \pm 0.7 \pm 0.6 \mevcc\,. 
\end{equation*}
The~measured mass agrees well 
 with the~value 
%% \begin{equation*}
%%  \upmu_{\Ppsi\mathrm{(3770)}}  = 3779.8 \pm 0.6 \mevcc\,, 
%% \end{equation*}
determined 
by Shamov and Todyshev~\cite{Shamov:2016mxe}
from available \mbox{$\mathrm{e}^+\mathrm{e}^-$}~cross\nobreakdash-section data.
It also agrees
well with and has a~better precision than the~current 
world average~\cite{PDG2018}, referred as PDG average in Table~\ref{tab:results:psi},
%% \begin{equation*}
%%   \upmu_{\Ppsi\mathrm{(3770)}}  = 3778.1 \pm 1.2 \mevcc\,,
%% \end{equation*}
which is dominated by the~value measured by the~KEDR collaboration~\cite{Anashin:2011kq}.
Reference~\cite{PDG2018} also quotes a~value,
referred as PDG fit,
resulting from 
a~fit that includes precision measurements of the~mass 
difference between the~$\Ppsi\mathrm{(3770)}$ and 
$\Ppsi\mathrm{(2S)}$ states made by 
the~BES collaboration~\cite{Ablikim:2007gd,Ablikim:2006md,Ablikim:2006zq}.
%% \begin{equation*}
%% \upmu_{\Ppsi\mathrm{(3770)}}  = 3773.13 \pm 0.35 \mevcc\,. 
%% \end{equation*}
Both~the~measurement made here and the~PDG average are in disagreement with the~PDG fit value. 

\begin{table}[tb]
  \centering
  \caption{ \small
    Summary of mass measurements for the~$\Ppsi\mathrm{(3770)}$~state.
  }\label{tab:results:psi}
  \vspace*{3mm}
  \begin{tabular*}{0.70\textwidth}{@{\hspace{5mm}}l@{\extracolsep{\fill}}cl@{\hspace{5mm}}}
    & 
    & $~~m_{\Ppsi\mathrm{(3770)}}~~\left[\!\mevcc\right]$
    \\[1mm]
    \hline 
    \\[-2mm]
        Shamov and Todyshev              
     &  \cite{Shamov:2016mxe}
     &  $3779.8\phantom{0} \pm 0.6\phantom{0}\phantom{\pm0.00}$
     \\
     PDG~average            
     &   \cite{PDG2018}        
     &   $3778.1\phantom{0} \pm 1.2\phantom{0}\phantom{\pm0.00}$
     \\
      PDG fit                 
     & \cite{PDG2018} 
     & $3773.13 \pm 0.35\phantom{\pm0.00}$
     \\
    %%This analysis &    &  $3778.13 \pm 0.70 \pm 0.63$
     This analysis &    &  $3778.1\phantom{3}\pm 0.7 \pm 0.6\phantom{3}$
  \end{tabular*}
\end{table}

%\clearpage

% Do not include this in any draft (just for information in the template)
%\input{acknowledgements_intro}
% Comment this in for paper drafts; do not include this in analysis note and conference reports
\section*{Acknowledgements}
%
% These Acknowledgements valid from 14-Aug-2018
%
\noindent We express our gratitude to our colleagues in the~CERN
accelerator departments for the~excellent performance of the~LHC. 
We~thank the technical and administrative staff at the~LHCb
institutes.
We~acknowledge support from CERN and from the national agencies:
CAPES, CNPq, FAPERJ and FINEP\,(Brazil); 
MOST and NSFC\,(China); 
CNRS/IN2P3\,(France); 
BMBF, DFG and MPG\,(Germany); 
INFN\,(Italy); 
NWO\,(Netherlands); 
MNiSW and NCN\,(Poland); 
MEN/IFA\,(Romania); 
MSHE\,(Russia); 
MinECo\,(Spain); 
SNSF and SER\,(Switzerland); 
NASU\,(Ukraine); 
STFC\,(United Kingdom); 
NSF\,(USA).
We~acknowledge the computing resources that are provided by CERN, 
IN2P3\,(France), KIT and DESY\,(Germany), INFN\,(Italy), 
SURF\,(Netherlands),
PIC\,(Spain), GridPP\,(United Kingdom), RRCKI and Yandex
LLC\,(Russia), CSCS\,(Switzerland), 
IFIN\nobreakdash-HH\,(Romania), CBPF\,(Brazil),
PL\nobreakdash-GRID\,(Poland) and OSC\,(USA).
We~are indebted to the communities behind the~multiple open-source
software packages on which we depend.
Individual groups or members have received support from
AvH Foundation (Germany);
EPLANET, Marie Sk\l{}odowska\nobreakdash-Curie Actions and ERC\,(European Union);
ANR, Labex P2IO and OCEVU, and R\'{e}gion Auvergne-Rh\^{o}ne-Alpes\,(France);
Key Research Program of Frontier Sciences of CAS, CAS PIFI, 
and the Thousand Talents Program\,(China);
RFBR, RSF and Yandex LLC\,(Russia);
GVA, XuntaGal and GENCAT\,(Spain);
the Royal Society
and the Leverhulme Trust (United Kingdom);
Laboratory Directed Research and Development program of LANL\,(USA).

% This should be taken out in the final paper
\clearpage

\addcontentsline{toc}{section}{References}
%\setboolean{inbibliography}{true}
\bibliographystyle{LHCb}
\bibliography{local,main,standard,LHCb-PAPER,LHCb-CONF,LHCb-DP,LHCb-TDR}

\newpage
% LHCb Collaboration author list
% Data extracted on March 25th, 2019 at 3:35pm for reference date 07-Feb-2019
\centerline
{\large\bf LHCb Collaboration}
\begin
{flushleft}
\small
R.~Aaij$^{29}$,
C.~Abell{\'a}n~Beteta$^{46}$,
B.~Adeva$^{43}$,
M.~Adinolfi$^{50}$,
C.A.~Aidala$^{77}$,
Z.~Ajaltouni$^{7}$,
S.~Akar$^{61}$,
P.~Albicocco$^{20}$,
J.~Albrecht$^{12}$,
F.~Alessio$^{44}$
% ...

{\footnotesize \it

% ...
$ ^{7}$Universit{\'e} Clermont Auvergne, CNRS/IN2P3, LPC, Clermont-Ferrand, France\\
% ...
$ ^{12}$Fakult{\"a}t Physik, Technische Universit{\"a}t Dortmund, Dortmund, Germany\\
% ...
$ ^{20}$INFN Laboratori Nazionali di Frascati, Frascati, Italy\\
% ...
$ ^{29}$Nikhef National Institute for Subatomic Physics, Amsterdam, Netherlands\\
% ...
$ ^{43}$Instituto Galego de F{\'\i}sica de Altas Enerx{\'\i}as (IGFAE), Universidade de Santiago de Compostela, Santiago de Compostela, Spain\\
$ ^{44}$European Organization for Nuclear Research (CERN), Geneva, Switzerland\\
% ...
$ ^{46}$Physik-Institut, Universit{\"a}t Z{\"u}rich, Z{\"u}rich, Switzerland\\
% ...
$ ^{50}$H.H. Wills Physics Laboratory, University of Bristol, Bristol, United Kingdom\\
% ...
$ ^{61}$University of Cincinnati, Cincinnati, OH, United States\\
% ...
$ ^{63}$Syracuse University, Syracuse, NY, United States\\
% ...
$ ^{77}$University of Michigan, Ann Arbor, United States, associated to $^{63}$\\
% ...
}
\end{flushleft}

\end{document}

%%%%%%%%%%%%%%%%%%%%%%%%%
%%%%% Title     %%%%%%%%%
%%%%%%%%%%%%%%%%%%%%%%%%%
\renewcommand{\thefootnote}{\fnsymbol{footnote}}
\setcounter{footnote}{1}

% %%%%%%% CHOOSE TITLE PAGE--------
%\onecolumn
%\input{title-LHCb-INT}
%\input{title-LHCb-ANA}
%\input{title-LHCb-CONF}
%\input{title-LHCb-PAPER}
%\twocolumn
% %%%%%%%%%%%%% ---------

\renewcommand{\thefootnote}{\arabic{footnote}}
\setcounter{footnote}{0}

%%%%%%%%%%%%%%%%%%%%%%%%%%%%%%%%
%%%%%  Table of Content   %%%%%%
%%%%%%%%%%%%%%%%%%%%%%%%%%%%%%%%
%%%% Uncomment next 2 lines if desired
%\tableofcontents
%\cleardoublepage

%%%%%%%%%%%%%%%%%%%%%%%%%
%%%%% Main text %%%%%%%%%
%%%%%%%%%%%%%%%%%%%%%%%%%

\pagestyle{plain} % restore page numbers for the main text
\setcounter{page}{1}
\pagenumbering{arabic}

%% Uncomment during review phase. 
%% Comment before a final submission.
%% \linenumbers

% You can include short sections directly in the main tex file.
% However, for larger papers it is desirable to split the text into
% several semiautonomous files, which can be revised independently.
% This is especially useful when developing a document in
% collaboration with several people, since then different parts can be
% edited independently.  This type of file organization is shown here.
%

%% \input{supplementary-app}

\section*{Supplementary material for LHCb-PAPER-2019-005}
\label{sec:Supplementary-App}

This appendix contains supplementary material that will be posted on 
the~public CDS record but will not appear in the~paper.

\subsection*{Additional information on the~{\boldmath{$\Pchi_{\cquark2}\mathrm{(3930)}$}} state}
\begin{figure}[htb]
\begin{center}
\resizebox{5.7in}{!}{\includegraphics{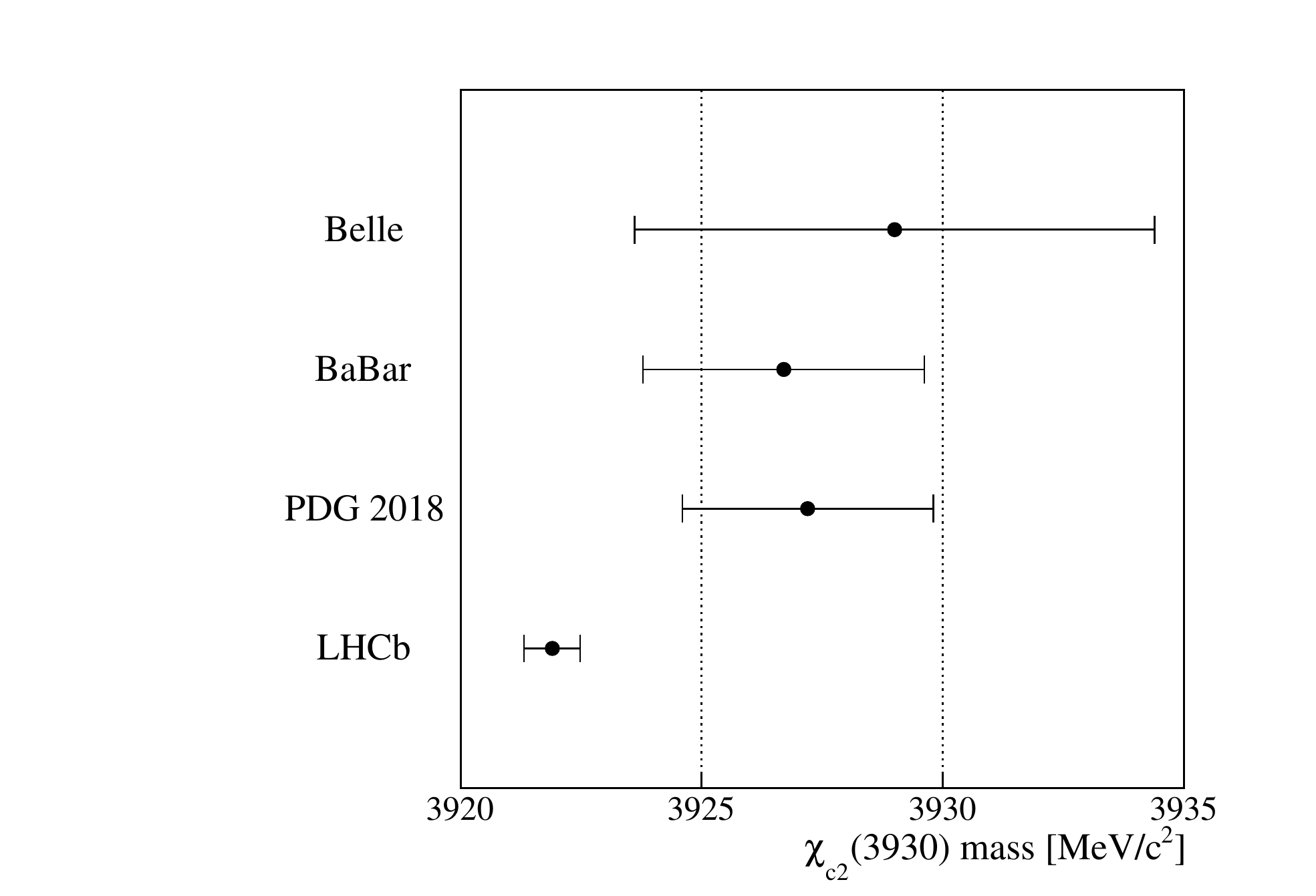}}
\resizebox{5.7in}{!}{\includegraphics{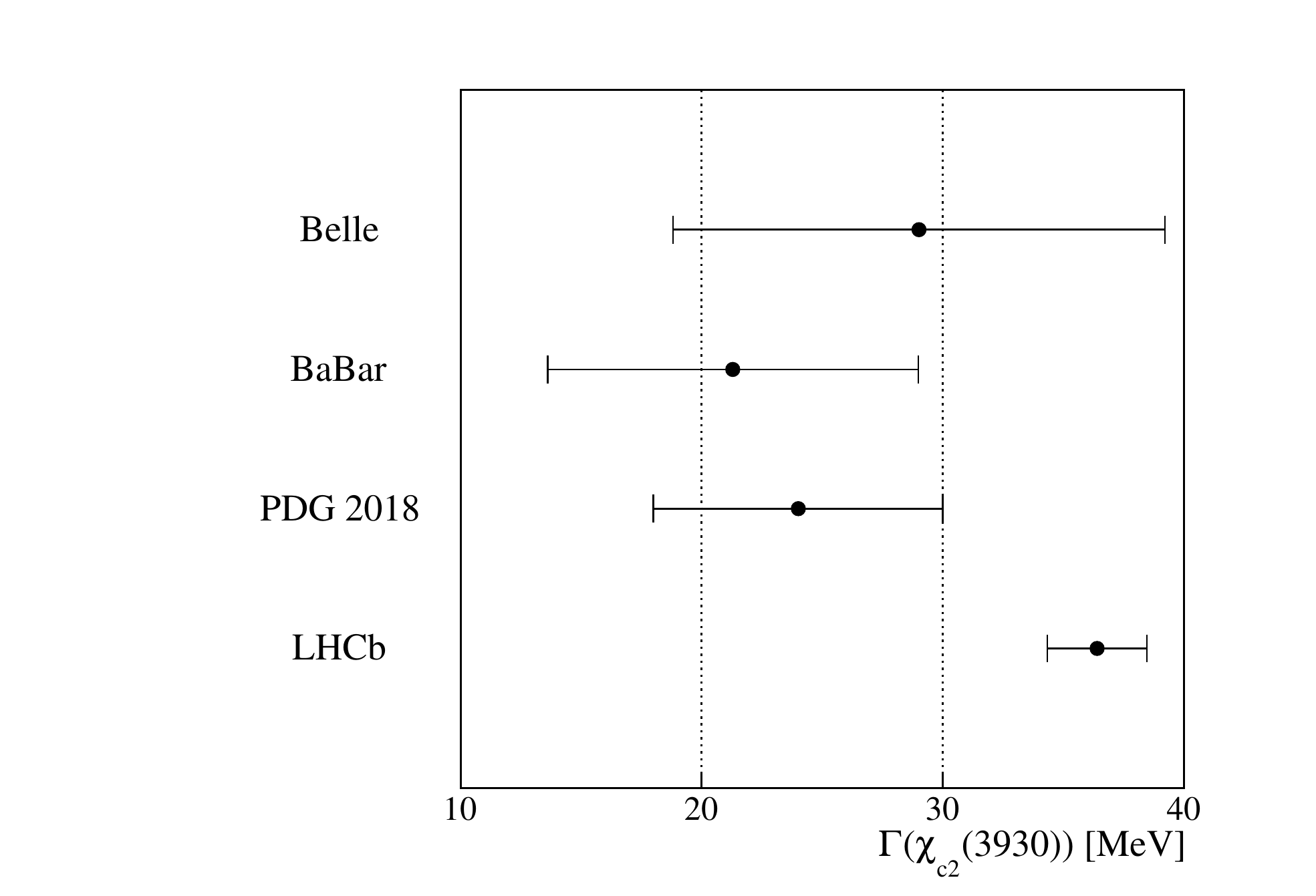}}
%\vspace{-5mm}
\caption{\small Measurements of the $\Pchi_{\cquark2}\mathrm{(3930)}$ (top)~mass 
and (bottom)~width by the
  Belle \cite{Uehara:2005qd} and BaBar \cite{Aubert:2010ab} collaborations
  together with the average calculated by the PDG \cite{PDG2018} and
  the LHCb measurement.}
\label{fig:chic2mass}
\end{center}
\end{figure}

The $\Pchi_{\cquark2}\mathrm{(3930)}$ meson was first observed by the B-factories \cite{Uehara:2005qd, Aubert:2010ab} in the reaction \decay{\gamma\gamma}{\D \Db}. The LHCb results for the mass and natural width of this resonance are compared to the B-factory values in Fig.~\ref{fig:chic2mass}.
The mass measured here is $2 \sigma$ lower than the world average whilst the natural width is $2 \sigma$ higher.

The~B\nobreakdash-factories
  also reported evidence for a~second state, the~$\mathrm{X(3915)}$,
   that decays to
   $\jpsi\upomega$~\cite{Abe:2004zs,Uehara:2009tx,delAmoSanchez:2010jr,Lees:2012xs}. The Review of Particle Properties \cite{PDG2018} gives the mass of this state as 
  \begin{equation*}
  m_{\mathrm{X(3915)}} = 3918.4 \pm 1.9 \mevcc\,. 
\end{equation*} 
Based upon an analysis of one-dimensional angular distributions~\cite{Lees:2012xs} and the assumption that a~$J^{PC}=2^{++}$~state is produced only with helicity $\pm 2$, as is expected for a pure charmonium state, the $\mathrm{X(3915)}$ was assigned spin-parity $0^{++}$. A natural interpretation would then be that it is the $\Pchi_{\cquark0}\mathrm{(2P)}$~state. However, as discussed in Refs.~\cite{Guo:2012tv,Olsen:2014maa,Zhou:2015uva}, this assignment is problematic since the~natural width of the~$\Pchi_{\cquark0}\mathrm{(2P)}$~state is expected to be larger. In~addition, the~$\Pchi_{\cquark0}\mathrm{(2P)}$~state should have a~large branching fraction to $\D\Dbar$~final state whereas there is no evidence  for the~$\mathrm{X(3915)}$~state decaying to open charm. Reference~\cite{Zhou:2015uva} proposes that
  the~$\mathrm{X(3915)}$ and the $\Pchi_{\cquark2}\mathrm{(3930)}$ states are the~same state with spin\nobreakdash-parity assignment $2^{++}$. This requires that 
  the~zero\nobreakdash-helicity amplitude dominates due to a significant non-$\cquark\overline{\cquark}$ contribution to the~wave function.
  The~Belle collaboration has subsequently observed another~state,
  $\Pchi_{\cquark0}\mathrm{(3860)}$, which has a
  large natural width and decays to $\D\Db$~final state. 
  This is a better candidate to be the~$\Pchi_{\cquark0}\mathrm{(2P)}$~state~\cite{Chilikin:2017evr}.
  The~question of the~nature and existence of the~$\mathrm{X(3915)}$~state remains open. It is interesting to note that the value of the mass measured here is roughly midway between the~values the~PDG quotes for the $\Pchi_{\cquark2}\mathrm{(3930)}$ and the~$\mathrm{X(3915)}$~states. Further studies are needed to understand if there are one or two distinct charmonium states in this region.

\subsection*{Additional information on the {\boldmath{$\Ppsi\mathrm{(3770)}$}} mass}
\begin{figure}[t]
\begin{center}
\resizebox{5.5in}{!}{\includegraphics{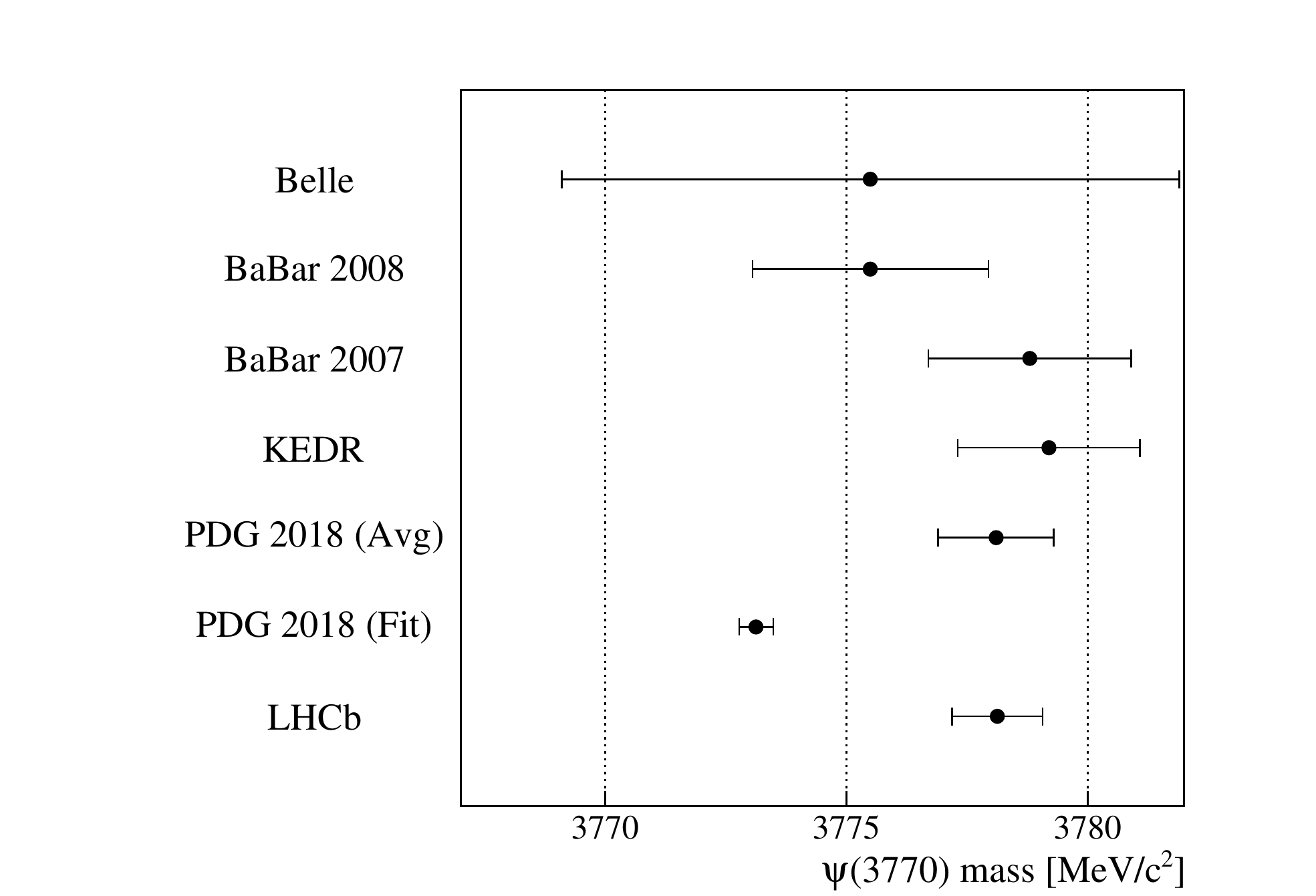}}
%\vspace{-5mm}
\caption{\small Measurements of the $\Ppsi\mathrm{(3770)}$ mass by the
  Belle \cite{Brodzicka:2007aa}, BaBar \cite{Aubert:2007rva,
    Aubert:2006mi} and KEDR \cite{Anashin:2011kq} collaborations
  together with the average calculated by the PDG \cite{PDG2018} and
  the LHCb measurement. The measurements are ordered according to decreasing total
uncertainty, which is the sum of statistical and systematic
uncertainties in quadrature. The PDG fit value is also shown.}
\label{fig:psimass}
\end{center}
\end{figure}

Figure \ref{fig:psimass} summarises the measurements of the
$\Ppsi\mathrm{(3770)}$ mass used by the~PDG to calculate its average. Our measurement is in good agreement. The PDG average does not include the BES-II measurement~\cite{Ablikim:2007gd,Ablikim:2006md,Ablikim:2006zq},
\begin{equation*}
  \m_{\Ppsi\mathrm{(3770)}}  = 3772.0 \pm 1.9 \mevcc\,,
\end{equation*}
given in Ref.\cite{Ablikim:2007gd} since it does not include the effect of
interference between resonant and non-resonant $\D\Db$ production. The
PDG average and our measurement also agree with the analysis of available 
$\mathrm{e}^+\mathrm{e}^-$
cross-section data in Ref.~\cite{Shamov:2016mxe}
\begin{equation*}
  m_{\Ppsi\mathrm{(3770)}}  = 3779.8 \pm 0.6 \mevcc\,. 
\end{equation*}
The~PDG also quotes a~fit value that includes precision measurements of the mass difference between the $\Ppsi\mathrm{(3770)}$ and $\Ppsi\mathrm{(2S)}$ states made by the~BES collaboration~\cite{Ablikim:2007gd,Ablikim:2006md,Ablikim:2006zq}.
\begin{equation*}
  m_{\Ppsi\mathrm{(3770)}}  = 3773.13 \pm 0.35 \mevcc\,. 
\end{equation*}
%% Both the measurement made here and the PDG average are in disagreement with the PDG fit value. 

\clearpage 
\addcontentsline{toc}{section}{References}
%\setboolean{inbibliography}{true}
\bibliographystyle{LHCb}
\bibliography{local,main,standard,LHCb-PAPER,LHCb-CONF,LHCb-DP,LHCb-TDR}